\documentclass[amsmath,amssymb,aps,prb,twocolumn]{revtex4-2}
\usepackage{graphicx}
\usepackage{bm}
\usepackage{array}
\usepackage{amsmath}
\usepackage{txfonts}
\usepackage{color}
\usepackage{booktabs}

\begin{document}


\title{Ultralow temperature NMR of CeCoIn$_5$}

\author{M. Yamashita$^1$}
 \email[]{my@issp.u-tokyo.ac.jp}
\author{M. Tashiro$^1$}
\author{K. Saiki$^1$}
\author{S. Yamada$^1$}
\author{M. Akazawa$^1$}
\author{M. Shimozawa$^{1,2}$}
\author{T. Taniguchi$^{1,3}$}
\author{H. Takeda$^1$}
\author{M. Takigawa$^1$}
\author{H. Shishido$^{4,5}$}

\affiliation{$^1$The Institute for Solid State Physics, The University of Tokyo, Kashiwa, 277-8581, Japan}
\affiliation{$^2$Graduate School of Engineering Science, Osaka University, Toyonaka, 560-8531, Japan}
\affiliation{$^3$Institute for Materials Research, Tohoku University, Sendai 980-8577, Japan}
\affiliation{$^4$Department of Physics and Electronics, Graduate School of Engineering, Osaka Prefecture University, Sakai, Osaka 599-8531, Japan.}
\affiliation{$^5$Institute for Nanofabrication Research, Osaka Prefecture University, Sakai, Osaka 599-8531, Japan.}

\date{\today}

\begin{abstract}
We have performed $^{59}$Co NMR measurements of CeCoIn$_5$ down to Ultralow temperatures. We find that the temperature dependence of the spin-echo intensity provides a good measure of the sample temperature, enabling us to determine a pulse condition not heating up the sample by the NMR pulses down to ultralow temperatures. From the longitudinal relaxation time ($T_1$) measurements at 5\,T applied along the $c$ axis, a pronounced peak in $1/T_1T$ is observed at 20\,mK, implying an appearance of magnetic order as suggested by the recent quantum oscillation measurements [H. Shishido {\it et al.}, Phys. Rev. Lett. {\bf 120}, 177201 (2018)]. On the other hand, the NMR spectrum shows no change below 20\,mK. Moreover, the peak in $1/T_1 T$ disappears at 6 and 8\,T in contrast to the results of the quantum oscillation. We discuss that an antiferromagnetic state with a moment lying in the $a$--$b$ plane can be a possible origin for the peak in $1/T_1 T$ at 5\,T.
\end{abstract}

\maketitle

\section{INTRODUCTION}

Understanding the role of enhanced quantum fluctuations near a magnetic quantum critical point (QCP) has been a central issue in condensed-matter physics~\cite{Sachdev2010,Shibauchi2014}, because such quantum fluctuations are believed to mediate various exotic phenomena, such as non-Fermi liquid behaviors, enhancements of effective mass and unconventional Cooper pairings. Many studies of unconventional superconductivity near a magnetic QCP have been performed in heavy-fermion materials~\cite{Gegenwart2008}. The energy scale of the heavy-electron materials is reduced by the strong mass renormalization owing to hybridization of $f$-electrons with conduction electrons. This reduced energy scale enables a fine tuning of the system near a QCP by easily accessible magnetic fields or pressures, making heavy-electron materials as ideal platforms to study the QCP physics.

Among the various heavy-fermion materials, CeCoIn$_5$ has been attracting broad attention because of its $d$-wave superconducting state with the high transition temperature of 2.3\,K and the proximity to a putative magnetic QCP~\cite{Izawa2001,Kawasaki2003,Tokiwa2013}. A field-induced QCP has been inferred to lie near the upper critical field from the crossover of non-Fermi liquid behaviors at low fields to Fermi liquid behaviors at higher fields~\cite{Bianchi2003,Paglione2003,Howald2011,Zaum2011}. However, no antiferromagnetic (AFM) state corresponding to the QCP has been observed, shrouding the origin of the QCP in mystery. It has been assumed that the apparent absence of AFM order can be explained by the AFM state being hidden at an inaccessible negative pressure~\cite{SarraoThompson2007} or superseded by the superconductivity~\cite{Tokiwa2013}.

Recently, some of us have reported anomalous decrease in the de Haas-van Alphen (dHvA) amplitudes of CeCoIn$_5$ below 20\,mK for 6--10\,T ($H \parallel c$) (Ref.\,\onlinecite{Shishido2018}).
An appearance of a field-induced AFM phase has been put forward to explain the decrease of the dHvA amplitude, because additional dissipation by magnetic breakdowns in the AFM phase can provide the most plausible explanation for the decrease of the dHvA amplitude.
However, direct evidence of magnetic order has yet to be observed.

It is a very challenging issue to investigate a presence of magnetic order below 20\,mK, unattainable low temperatures for a conventional dilution refrigerator, under high fields of 6--10\,T.
One of the most powerful methods to elucidate magnetic order is a neutron-diffraction measurement. 
Neutron-diffraction experiments were performed down to nanokelvin and picokelvin temperatures at Ris{\o} National Laboratory in Denmark to find nuclear magnetic order in simple metals~\cite{OjaLounasmaa1997}, which has been shutdown. 
SQUID-based magnetization measurements are also powerful probes for detecting nuclear magnetism~\cite{OjaLounasmaa1997} and superconductivity~\cite{Buchal1983,Tuoriniemi2007, Schuberth2016, Prakash2017}  at ultralow temperatures, which, however, cannot be applied to high-field measurements required for CeCoIn$_5$.
Another candidate is a nuclear magnetic resonance (NMR) measurement, which is sensitive to changes of the internal magnetic fields and the spin dynamics.
 It is, however, necessary to pay attention to a sample heating caused by the NMR pulses~\cite{Pustogow2019,Ishida2020}.

In this paper, to find evidence of the magnetic order in CeCoIn$_5$, we have utilized the spin-echo NMR technique down to ultralow temperatures.
We find that the repetition time dependence and the temperature dependence of the spin-echo intensity allow one to find pulse conditions in which the heating of the sample is kept negligibly small.
By this pulse condition, we have investigated the temperature dependence of the longitudinal relaxation time ($T_1$) of $^{59}$Co NMR.
We find that $1/T_1 T$ shows a peak near 20\,mK at 5\,T. However, this peak is absent at 6 and 8\,T. Moreover, the NMR spectrum at 5\,T shows no change below 20\,mK. We discuss that magnetic order with a moment lying in the $a$--$b$ plane is one possible origin.

\section{EXPERIMENTAL DETAILS}

\subsection{Materials and methods}

\begin{figure}[!tbh]
	\centering
	\includegraphics[width=0.8\linewidth]{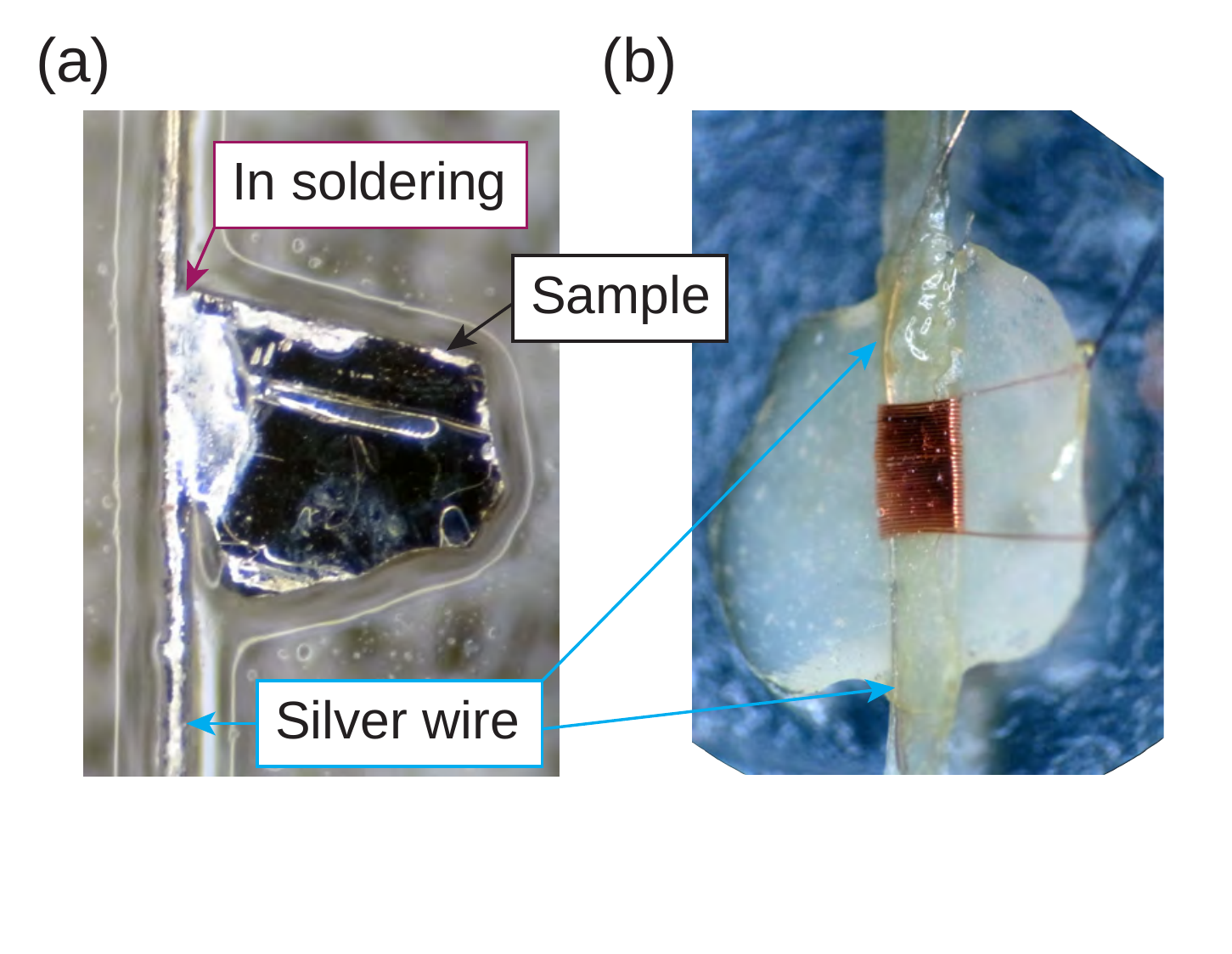}
	\caption{
		(a) A picture of the sample with a silver wire (100\,$\mu$m diameter) soldered by indium. The sample size is 1.1\,mm {$\times$} 1.0\,mm {$\times$} 0.1\,mm (b) Cu wires were wounded around the sample for NMR measurements. The silver wire was thermally anchored to the cryostat.
	}
	\label{crypic}
\end{figure}

High-quality single crystals of CeCoIn$_5$ were grown by the In-flux method~\cite{Shishido2002}. NMR measurements were performed under a magnetic field $H \parallel c$ down to 5\,mK using our homemade nuclear-demagnetization cryostat. To ensure the sample temperature down to the ultralow temperature, the sample and the NMR pickup coil were immersed in liquid $^3$He. For further thermal anchoring, a silver wire of 100\,$\mu$m diameter was soldered to the sample by indium (Fig.\,\ref{crypic} (a) and (b)). This silver wire was then firmly anchored to the cryostat. The cryostat temperature was measured by a melting curve thermometer as described in Ref.~\onlinecite{Shishido2018}.

NMR measurements were performed by the standard spin-echo method. The first and the second pulses with a common amplitude of $V_{\textrm{pls}}$ and a pulse width of $t_{\textrm{1st}}$ and $t_{\textrm{2nd}} $ ($= 2 \times t_{\textrm{1st}}$) were applied with the interval time $\tau$ between the first and the second pulses to observe a spin-echo signal (details of the NMR sequence are described in Appedix\,\ref{sec:t_rp_dep}).
While the spin-echo intensity is proportional to $V_{\textrm{pls}} (t_{\textrm{1st}} + t_{\textrm{2nd}})$, the pulse power ($P_{\textrm{pls}}$) applied to the NMR circuit is proportional to $V_{\textrm{pls}}^2 (t_{\textrm{1st}} + t_{\textrm{2nd}})$
(in this paper, we determined  $V_{\textrm{pls}}$ as the input voltage applied to the NMR tune circuit, and define $P_{\textrm{pls}} = V_{\textrm{pls}}^2 (t_{\textrm{1st}} + t_{\textrm{2nd}})$ for the sake of expedience).
Therefore, the heating by the NMR pulses can be effectively suppressed by reducing $V_{\textrm{pls}}$ with keeping $V_{\textrm{pls}} (t_{\textrm{1st}} + t_{\textrm{2nd}})$ with a longer pulse.
A typical pulse width used below 100\,mK was $\sim$100\,$\mu$s, and the tipping angle was 20--30\,degree.
The amplitude and the width of the comb pulse used for the $T_1$ measurements was the same with those of the first pulse.

\begin{figure*}[!tbh]
	\centering
	\includegraphics[width=0.8\linewidth]{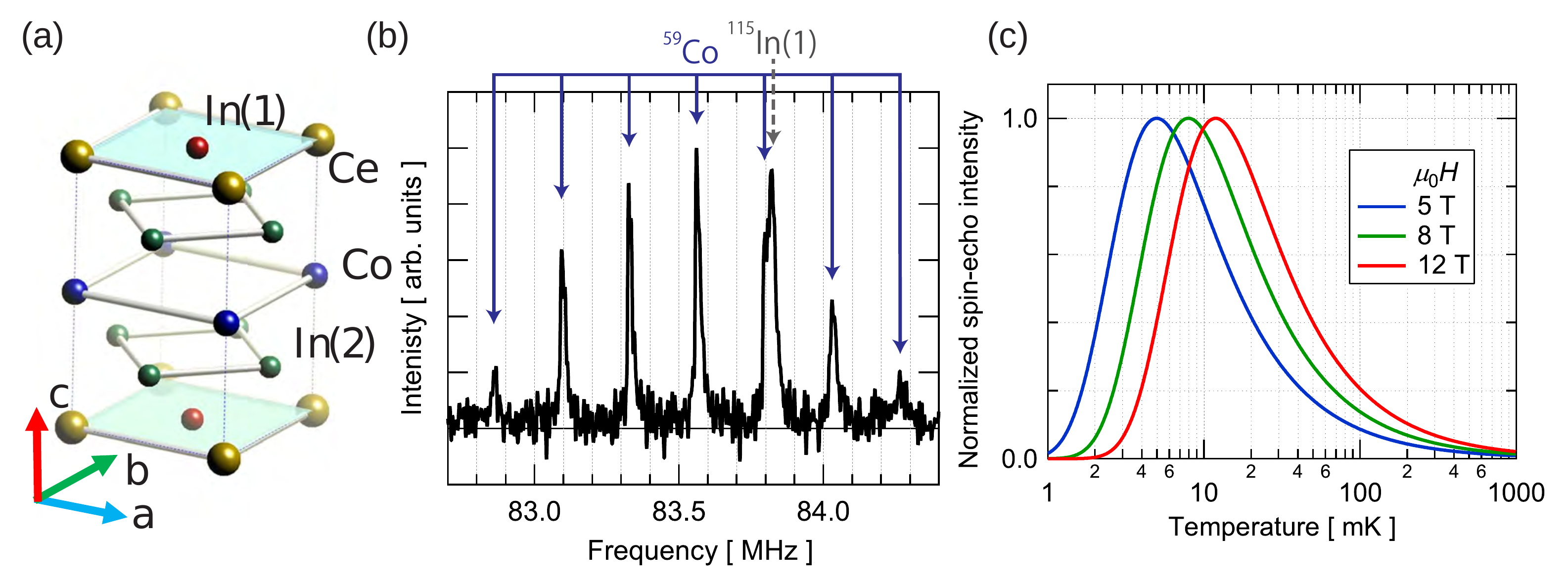}
	\caption{
		(a) Crystal structure of CeCoIn$_5$. (b) NMR spectrum taken at 1.7\,K and 8\,T ($H \parallel c$). Seven absorption peaks corresponding to $^{59}$Co ($I = 7/2$) were clearly observed. Note that a $^{115}$In NMR signal from the In(1) site is merged with the first quadrupole-satellite peak at the higher-frequency side. (c) The calculated temperature dependence of the spin-echo intensity of the center line at different fields. Each curve is normalized by the peak value.
	}
	\label{NMRspct}
\end{figure*}

In CeCoIn$_5$, NMR measurements can be performed at $^{59}$Co ($I = 7/2$) and $^{115}$In ($I = 9/2$) nuclei (Fig.\,\ref{NMRspct}(a)). Figure\,2(b) shows a NMR spectrum at 1.7\,K and 8\,T. The NMR spectra were obtained by summing the Fourier transform of the spin-echo signal measured at equally spaced rf-frequencies at a fixed magnetic field. Seven absorption peaks of $^{59}$Co  ($I = 7/2$) were observed with an equal-frequency spacing. A NMR signal of $^{115}$In from one of the two inequivalent In sites, In(1),~\cite{Kumagai2011} was also observed.
In this work, we investigated the temperature dependence of the NMR spectrum and that of $1/T_1 T$ by measuring the center peak of the $^{59}$Co  signal.

\subsection{Estimation of the sample temperature}

For NMR measurements at low temperatures, measuring the actual sample temperature is strikingly important, because the heating by excess $P_{\textrm{pls}}$ might increase only the sample temperature without affecting the cryostat temperature. The most direct method to check the sample temperature is to observe the pulse-power dependence of the Knight shift after the NMR pulse, as discussed in the recent measurements~\cite{Pustogow2019,Ishida2020} of Sr$_2$RuO$_4$. In CeCoIn$_5$, the temperature dependence of the Knight shift becomes smaller below 100\,mK,~\cite{Sakai2011} not allowing us to check the sample temperature by measuring the Knight shift at ultralow temperatures.

We thus utilize the temperature dependence of the spin-echo intensity. The spin-echo intensity of an absorption peak is determined by the population difference between the neighboring energy levels ($m = -7/2$ to +7/2 for $^{59}$Co). Owing to the Boltzmann distribution of the nuclear spins, the spin-echo intensity $I_{\textrm{SE}}$ for the center peak of $^{59}$Co is given by

\begin{eqnarray} \label{eq:I_SE}
I_{\textrm{SE}}(T) \propto \frac{\exp \left( -\frac{E_{1/2}}{k_B T} \right) - \exp \left( -\frac{E_{-1/2}}{k_B T} \right)}{\sum_{m} \exp \left( -\frac{E_m}{k_B T} \right)},
\end{eqnarray}
where $k_B$ is the Boltzmann constant,
\begin{eqnarray} \label{eq:E_m}
E_m = - m \gamma \hbar H_0 + \frac{1}{6}h \nu_Q
\left\{
3m^2 - \frac{7}{2} \left( \frac{7}{2} + 1 \right)
\right\}
\end{eqnarray}
the eigenvalue of the energy for $m$-th energy level, $H_0$ the static magnetic field applied along the $c$ axis, $\gamma/2\pi = 10.1021310$\,MHz/Tesla the gyromagnetic ratio, and $\nu_Q = 0.230$\,MHz the nuclear quadrupolar resonance (NQR) frequency of $^{59}$Co (Refs.~\onlinecite{Kohori2001,Curro2001}).

The calculated temperature dependence of $I_{\textrm{SE}} (T)$ (Fig.\,\ref{NMRspct}(c)) shows a peak at the temperature where the thermal broadening $k_B T$ is about equal to the Zeeman gap $\gamma h H_0$. The peak temperature is 5--10\,mK for 5--12\,T, demonstrating a good temperature sensitivity at the ultralow temperatures. Thus, by checking whether the temperature dependence of $I_{\textrm{SE}}$ follows the curve given by Eq.\,(1), one can find an appropriate pulse condition (the amplitudes and the widths of the NMR pulses) by which the heating of the sample is negligible .


\section{Verification of the sample temperature by the temperature dependence of spin-echo intensity} \label{sec:verification}

\begin{figure*}[!tbh]
	\centering
	\includegraphics[width=0.9\linewidth]{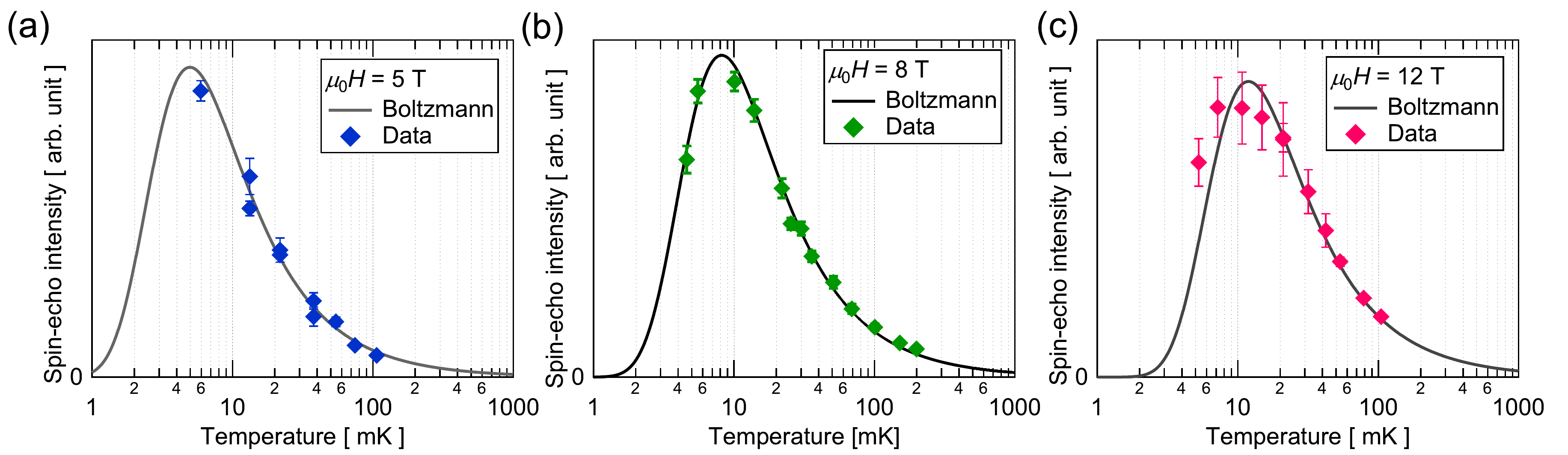}
	\caption{
		The temperature dependence of the spin-echo intensity at 5, 8, and 12\,T. The solid line shows the Boltzmann curve (Eq. (1)) for each temperature.
		The error bars represent $\pm \sigma$, where $\sigma$ is one standard deviation of the data.
		}
	\label{Boltzmann}
\end{figure*}

Here, we discuss the procedure to find a pulse condition with a negligible sample heating and how we confirmed that $I_{\textrm{SE}} (T)$ follows the curve shown in Fig.\,\ref{Boltzmann}(c).
We find that, because of the temperature dependence of $I_{\textrm{SE}}$, the repetition time ($t_{\textrm{rp}}$) dependence of $I_{\textrm{SE}}$ can be used to check the heating of the sample given by the pulse condition.
Thus, we determined the pulse condition at each temperature by checking the $t_{\textrm{rp}}$ dependence of $I_{\textrm{SE}}$ (see Appendix\,\ref{sec:t_rp_dep} for details).
We then measured the temperature dependence of the transverse relaxation time ($T_2$) to evaluate $I_{\textrm{SE}} (T)$ at $\tau=0$ for each temperature (see Appendix\,\ref{sec:T2}).
Finally, we took into account the tipping angle difference in the different temperature ranges.
In the pulse condition without the $t_{\textrm{rp}}$ dependence, the tipping angle was set as a smaller value in the lower temperature range.
Therefore, at the boundary temperature where the pulse condition was changed, the spin-echo signals were measured by both the tipping angles to normalize the signal intensities obtained by different tipping angles.

The temperature dependence of $I_{\textrm{SE}} (T)$ obtained by this procedure at 5, 8, and 12\,T is shown in Figs.\,\ref{Boltzmann}. As shown in Figs.\,\ref{Boltzmann}, the temperature dependence of $I_{\textrm{SE}} (T)$ observed by the pulse condition without the $t_{\textrm{rp}}$ dependence well follows the Boltzmann curve given by Eq.\,(1), ensuring the thermal equilibrium of the sample temperature with the cryostat temperature down to ultralow temperatures.
This good agreement, paving a way for reliable NMR measurements at ultralow temperatures, is one of the main outcomes of this work.

We note that, however, the validity of the thermal equilibrium is limited for a time scale longer than or equal to $T_1$ as discussed in detail below.
We also note that, as shown in Figs.\,\ref{Boltzmann}, the spin-echo intensity quickly decreases for $k_B T < \gamma h H_0$, because the nuclear spins are redistributed to lower-energy levels than those corresponding to the center peak of $^{59}$Co ($m = +1/2 \leftrightarrow -1/2$). 
At these low temperatures, the satellite peaks for lower-energy levels show larger signals (see Appendix\,\ref{sec:I_SE_rel} for details), offering alternative ways to extend measurements for lower temperatures at higher fields in future.

\section{Investigation of magnetic order in C{\MakeLowercase{e}}C{\MakeLowercase{o}}I{\MakeLowercase{n}}$_5$}

\subsection{Longitudinal relaxation time ($T_1$) measurements}

\begin{figure*}[!tbh]
	\centering
	\includegraphics[width=0.9\linewidth]{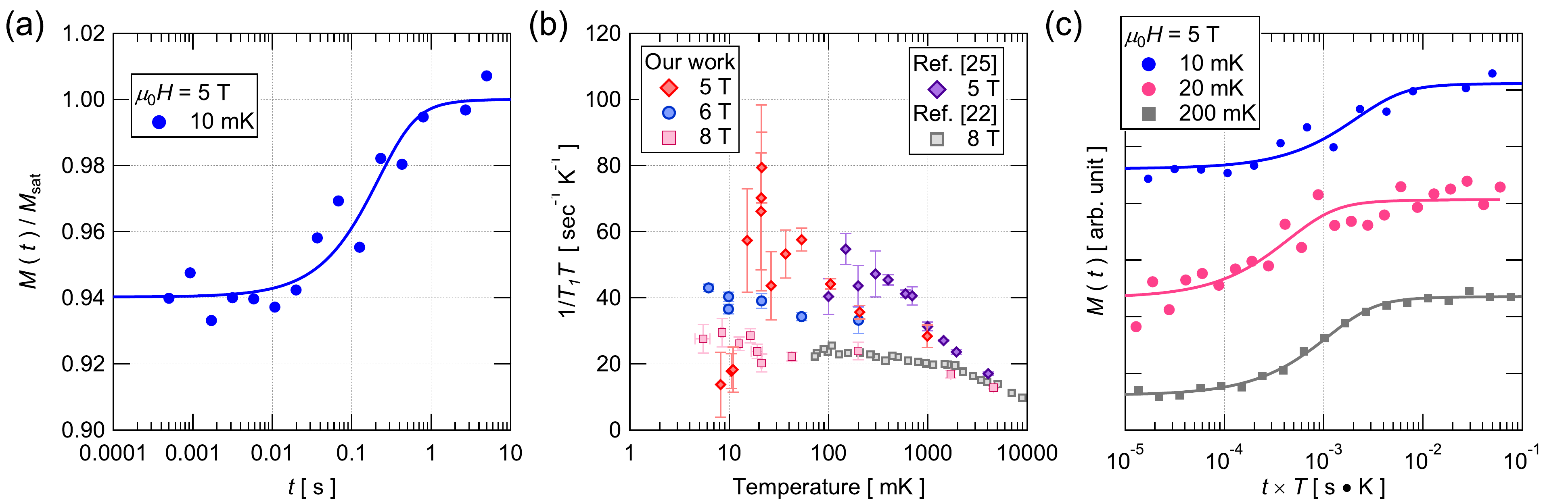}
	\caption{
		(a) Recovery curve of the magnetization $M(t)$ divided by the saturation value $M_{\textrm{sat}}$ at 5\,T and 10\,mK. The solid line shows a fit for the central transition of the nuclear spin 7/2.
		(b) The temperature dependence of $1/T_1 T$ at 5, 6, and 8\,T. The data of the previous work is taken from Ref.~\onlinecite{Taniguchi2020} (5\,T) and Ref.~\onlinecite{Sakai2011} (8\,T).
		The error bars represent $\pm \sigma$, where $\sigma$ is one standard deviation of the fitting of the recovery curve by Eq.\,\ref{T1eq}.
		To show the overlapped data points at 5\,T,  all the data at 5\,T is listed in Table\,\ref{T1table} in Appendix\,\ref{sec:T1table}.
		(c) Comparison of recovery curves at 5\,T at different temperatures.
		To compare $T_1 T$, rather than $T_1$, at different temperatures, the data is plotted as a function of the product of the delay time ($t$) and the temperature. 
	}
	\label{T1}
\end{figure*}

We performed the longitudinal-relaxation time ($T_1$) measurements at 5, 6, and 8\,T by the pulse condition determined as described in the previous section. A typical recovery curve at 5\,T is shown in Fig.\,\ref{T1}(a). As shown in the solid line in Fig.\,\ref{T1}(a), the data is well fitted by the relaxation curve expected for the central transition of the nuclear spin 7/2 \cite{Narath1967}
\begin{eqnarray}
\begin{split}
1 - \frac{M(t)}{M_\textrm{sat}} &\propto 
\frac{1}{42} e^{-t/T_1} + \frac{3}{22} e^{-6t/T_1} \\
&\quad + \frac{75}{182} e^{-15t/T_1} + \frac{1225}{858} e^{-28t/T_1}.
\end{split}
\label{T1eq}
\end{eqnarray}

We determine $T_1$ by using Eq.\,(\ref{T1eq}) for the recovery curves observed at different temperatures and plot the temperature dependence of $1/T_1 T$ in Fig.\,\ref{T1}(b). 
As show in Fig.\,\ref{T1}(b), we confirm a good reproducibility of our data above $\sim 100$\,mK~\cite{footnote1} with those in Ref.~\cite{Taniguchi2020, Sakai2011}.

We find that $1/T_1 T$ at 5\,T shows a peak $\sim$20\,mK, whereas those at 6 and 8\,T follow the Korringa law ($1/T_1 T \sim$ constant) without a peak. The peak of $1/T_1 T$ at 5\,T can also be clearly inferred from the comparison of the recovery curves at 10, 20 and 200\,mK shown in Fig.\,\ref{T1}(c) where the data is plotted as a function of the product of the delay time ($t$) and the temperature to compare $T_1 T$, rather than $T_1$, of these curves. As shown in Fig.\,\ref{T1}(c), the recovery as a function of $t \times T$ takes place more quickly  at 20\,mK than those at 10 and 200\,mK.

The bigger errors of $1/T_1 T$ near the peak might reflect a sharp change of $1/T_1 T$.
Owing to the small tipping angle set to avoid the heating, a very long averaging time was required in the $T_1$ measurements to resolve a small change of the NMR signal. Therefore, a small temperature instability can easily cause a large ambiguity because of this long accumulation. At 5\,T, although we could observe the spin-echo signal down to 5\,mK by the pulse condition without the heating, we could not resolve the relaxation of the NMR signal down to 5\,mK owing to the small tipping angle.


\subsection{NMR spectrum below 20 mK}

\begin{figure}[!tbh]
	\centering
	\includegraphics[width=0.8\linewidth]{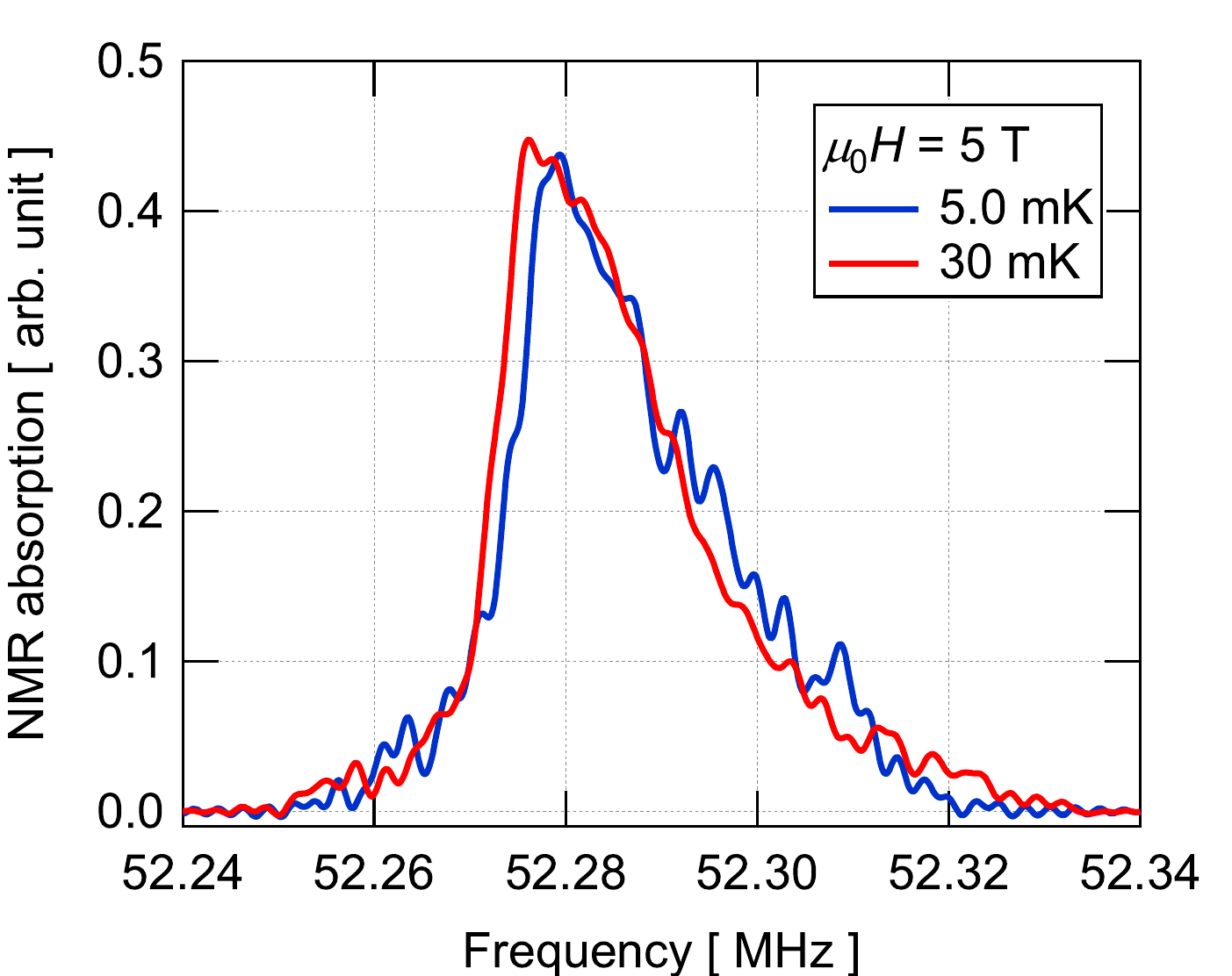}
	\caption{
		The NMR spectrum of $^{59}$Co center peak at 5.0 and 30\,mK.
	}
	\label{NMRspct20mK}
\end{figure}

To check if the peak of $1/T_1 T$ is caused by magnetic order, we compare the NMR spectrum above and below 20\,mK (Fig.\,\ref{NMRspct20mK}). As shown in Fig.\,\ref{NMRspct20mK}, the NMR spectrum at 5\,mK is virtually the same with that at 30\,mK.
From the line width of the NMR spectrum ($\sim$20\,kHz) and the coupling constant of $^{59}$Co (4.7\,kOe/$\mu_B$)~\cite{Kohori2001}, the absence of a broadening of the NMR spectrum limits the magnitude of the internal magnetic field along the applied field less than 0.004\,$\mu_B$.


\subsection{Possible magnetic order in CeCoIn$_5$ below 20\,mK}

In the dHvA measurements~\cite{Shishido2018}, the anomalous suppression of the dHvA amplitude has been observed at almost the constant temperature of $\sim$20\,mK above 7.5\,T. On the other hand, the peak in $1/T_1 T$ was observed only at 5\,T.
This field dependence may be attributed to a change of the $Q$ vector of the magnetic order. 

In general, $1/T_1T$ is given by
\begin{eqnarray}
\frac{1}{T_1 T} \propto \sum_{\bm q} f^2({\bm q}) \frac{{\textrm{Im\,}} \chi_{\perp}({\bm q},\omega_0)}{\omega_0},
\label{eqT1}
\end{eqnarray}
where $ f^2({\bm q})$ is the hyperfine form factor, $\textrm{Im\,} \chi_{\perp}({\bm q},\omega_0)$ is the imaginary part of the dynamical susceptibility perpendicular to the magnetic field, and $\omega_0 = 2\pi \gamma H_0$ is the Larmor frequency.
For the $^{59}$Co NMR in CeCoIn$_5$ under a magnetic field of $H \parallel c$, the hyperfine form factor $ f^2({\bm q})$ is written as 
\begin{eqnarray}
f^2({\bm q}) = 4 A_{\perp}^2 \cos^2 \pi q_c
\end{eqnarray}
owing to the  lattice symmetry of CeCoIn$_5$ (see Fig.\,\ref{NMRspct}(a)), where $A_{\perp}$ is the in-plane component of the hyperfine coupling constant~\cite{Sakai2010,Sakai2011} and $q_c$ is the $c$ component of the $Q$ vector.
This form factor filters out the spin fluctuations for a magnetic structure with $q_c=1/2$.
Therefore,  the presence and the absence of the  $1/T_1 T$ peak suggests a change of $q_c$ from an incommensurate $q_c \ne 1/2$ at 5\,T to $q_c = 1/2$ at higher fields.

We note that a similar change of the $Q$ vector has been observed in the related antiferromagnetic compound CeRhIn$_5$.
The magnetic structure of CeRhIn$_5$ at ambient pressure and at zero field~\cite{Bao2000} is
an incommensurate helical order with $Q_{\textrm{IC}}=(1/2,1/2,0.297)$.
This magnetic structure changes to a commensurate antiferromagnetic order with $Q_{\textrm C}=(1/2,1/2,1/2)$ under a high pressure above 1.7\,GPa (Ref.~\onlinecite{Yashima2009}), or to $Q'_{\textrm{C}}=(1/2,1/2,1/4)$ under a high magnetic field applied in the $a$--$b$ plane~\cite{Raymond2007}. 
This change of the $Q$ vector has been discussed in terms of the distance from the QCP~\cite{Yashima2009}.
Therefore, given that a field-induced QCP~\cite{Bianchi2003,Paglione2003,Howald2011,Zaum2011} has been discussed in CeCoIn$_5$ near $H_{\textrm{c2}}$ for $H \parallel c$, a change of the $Q$ vector can also be expected in CeCoIn$_5$.

Next, we discuss the absence of the change in the NMR spectrum below 20\,mK at 5\,T (Fig.\,\ref{NMRspct20mK}) despite the possible magnetic order.
The spectral shape corresponds to the histogram of the absolute values of the local magnetic field at the Co sites.  Because the local field is the vector sum of the external and the internal magnetic fields, appearance of the internal field parallel to the applied field ($\parallel c$ axis) results in a drastic change of the spectrum such as splitting or broadening.  On the other hand, a transverse internal field makes small change in the spectral shape.  
This is in contrast to that $1/T_1T$ probes the magnetic fluctuations perpendicular to the applied field (Eq.\,(\ref{eqT1})).  
Therefore, magnetic order only with transverse moments causes a peak in $1/T_1T$, but no change in the NMR spectrum.
In fact, such a magnetic structure has also been reported in CeRhIn$_5$, where the magnetic moment in the AFM phase is laid in the $a$--$b$ plane~\cite{Bao2000, Curro2000}.

\begin{figure}[!tbh]
	\centering
	\includegraphics[width=0.9\linewidth]{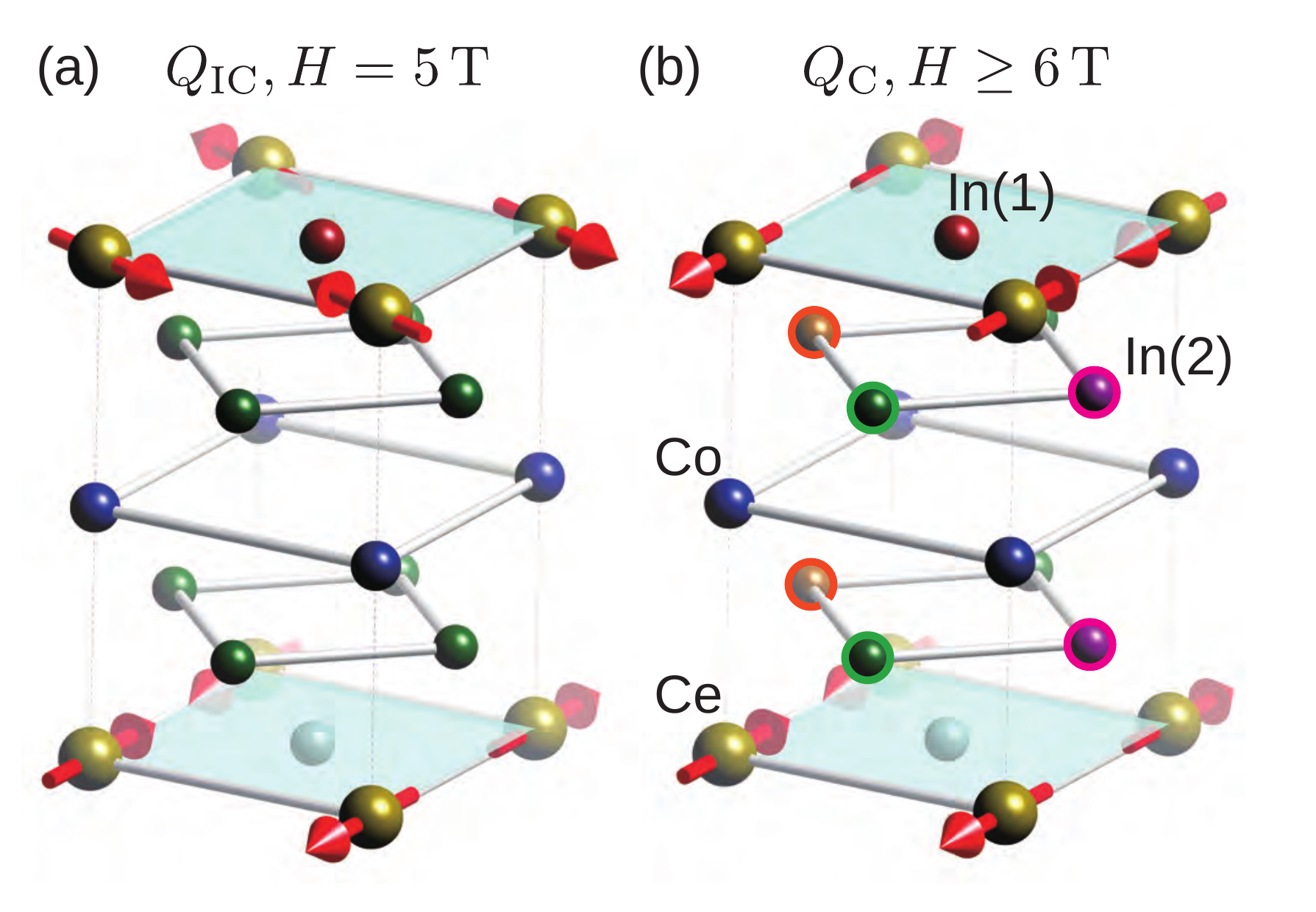}
	\caption{
		Proposed magnetic structures of CeCoIn$_5$ below 20\,mK with an incommensurate $Q_{\textrm{IC}}$ (a) and a commensurate $Q_{\textrm{C}}$ (b) by reference to that of CeRhIn$_5$ under pressure~\cite{Yashima2009}. The magnetic moments are suggested to be in-plane. Inequivalent In(2) sites for $Q_{\textrm C}$ are marked by solid circles with different colors in (b).
	}
	\label{MagStr}
\end{figure}

The possible magnetic structures of CeCoIn$_5$ are shown in Figs.\,\ref{MagStr}.
In the incommensurate magnetic order presumed at 5\,T (Fig.\,\ref{MagStr}(a)), the internal field at the Co site has only the in-plane component, which results in the anomaly in $1/T_1 T$ without the change of the NMR spectrum.
On the other hand, the internal field is completely canceled out owing to the local site symmetry at the Co site for the commensurate magnetic order presumed at higher fields (Fig.\,\ref{MagStr}(b)).
Although it is impossible to check the magnetic structure by using $^{59}$Co NMR, the $^{115}$In NMR measurements at the In(2) sites can be used to confirm the magnetic structure because of the lower site symmetry.
We have thus tried to perform $^{115}$In NMR measurements at the In(2) sites below 20\,mK.
However, as shown in Fig.\,\ref{T2}(a), $T_2$ at the In(2) sites is very short ($\sim 40$\,$\mu$s), disabling us to suppress the heating effect by a longer NMR pulse with a smaller amplitude.

It should be noted that the absence of the change in the NMR spectrum might be caused by a heating effect in the time scale shorter than $T_1$. The heating checks by the spin-echo intensity (the $t_{\textrm{rp}}$ dependence (Fig.\,\ref{RepDep}) and the temperature dependence of $I_{\textrm{SE}}$ (Fig.\,\ref{Boltzmann})) ensures the thermal equilibrium of the sample temperature in the time scale of $T_1$, but are insensitive to an instantaneous heating in the time scale of the Knight shift (0.1--1 ms, see Fig.\,\ref{PulseSeq}). The recent experiments~\cite{Pustogow2019,Ishida2020} checking the heating by the Knight shift of Sr$_2$RuO$_4$ report a short thermal relaxation of order of 1\,ms, which is much shorter than the delay time used for our $T_1$ measurements. Therefore, $T_{\textrm{ele}}$ might exceed 20\,mK in a time scale shorter than $\sim$1\,ms after the NMR pulses, but relax to below 20\,mK during the $T_1$ measurements. This explains the absence in the change of the NMR spectrum despite the appearance of a peak in $1/T_1 T$ by magnetic order. To check this possibility, we tried to observe the NMR spectrum by a weaker NMR power. However, the smaller NMR signal owing to the weaker NMR pulses disabled us to observe the NMR spectrum within our present resolution.
Improving the sensitivity of the NMR measurements is required to investigate a change in the NMR spectrum at 5\,T as well as to observe the $^{115}$In NMR at the In(2) sites down to ultralow temperatures, which remains as a future work.
Also, it is an important future issue to observe a broadening of the NMR spectrum under a tilted magnetic field to confirm the internal field in the AFM phase, although the magnitude of the magnetic moment might be comparable to the line width of the NMR spectrum (Fig.\,\ref{NMRspct20mK}) owing to the ultralow ordering temperature~\cite{fn}.

We note that another possible explanation for the $1/T_1 T$ peak, except for magnetic order, includes a crossover from a non-Fermi liquid to a Fermi liquid. As shown by the thermal expansion measurements~\cite{Zaum2011}, the crossover line is suggested to terminate near $H_{\textrm{c2}}$, which may correspond to 20\,mK at 5\,T.
However, this crossover should be observed at 6 and 8\,T at higher temperatures, which is absent in our measurements.


\section{Summary}

To investigate the origin of the anomaly found at $\sim$20\,mK in the quantum oscillation measurements~\cite{Shishido2018}, $^{59}$Co NMR measurements of CeCoIn$_5$ were performed down to 5\,mK under magnetic field of 5, 6, 8 and 12\,T applied parallel to the $c$ axis.

We have developed the spin-echo NMR measurements procedure to find a pulse condition without heating the sample down to ultralow temperatures. 
We find that, by tuning the pulse condition to avoid a repetition time dependence of the spin-echo intensity, the temperature dependence of the spin-echo intensity follows the curve expected by the Boltzmann distribution of the nuclear spins. This ensures no heating of the sample during the NMR measurements at least for the time scale of $T_1$ measurements. From the pulse condition determined by this procedure, we find a peak at $\sim$20\,mK in the temperature dependence of $1/T_1 T$ at 5\,T, whereas the peak is absent at 6 and 8\,T. The NMR spectrum at 5\,T shows no discernible change below 20\,mK.

We suggest that the presence and the absence of the peak in $1/T_1 T$ may be related to the change of the $Q$ vector in the AFM state from an incommensurate to a commensurate one as observed in CeRhIn$_5$~\cite{Yashima2009, Raymond2007}.
The absence of the change in the NMR spectrum may be explained by the magnetic moment lying only in the $a$--$b$ plane in the AFM state and/or a heating effect by the NMR pulse in the time scale much shorter than $T_1$.
Measurements of $^{115}$In NMR at the In(2) sites and $^{59}$Co NMR by weaker NMR pulses under a tilted magnetic field are remained as future works to reveal the magnetic ground state below 20\,mK.

\newpage


\appendix


\section{Optimizing pulse condition by repetition time ($t_{\textrm{rp}}$) dependence measurements} \label{sec:t_rp_dep}

\begin{figure*}[bth]
	\centering
	\includegraphics[width=0.8\linewidth]{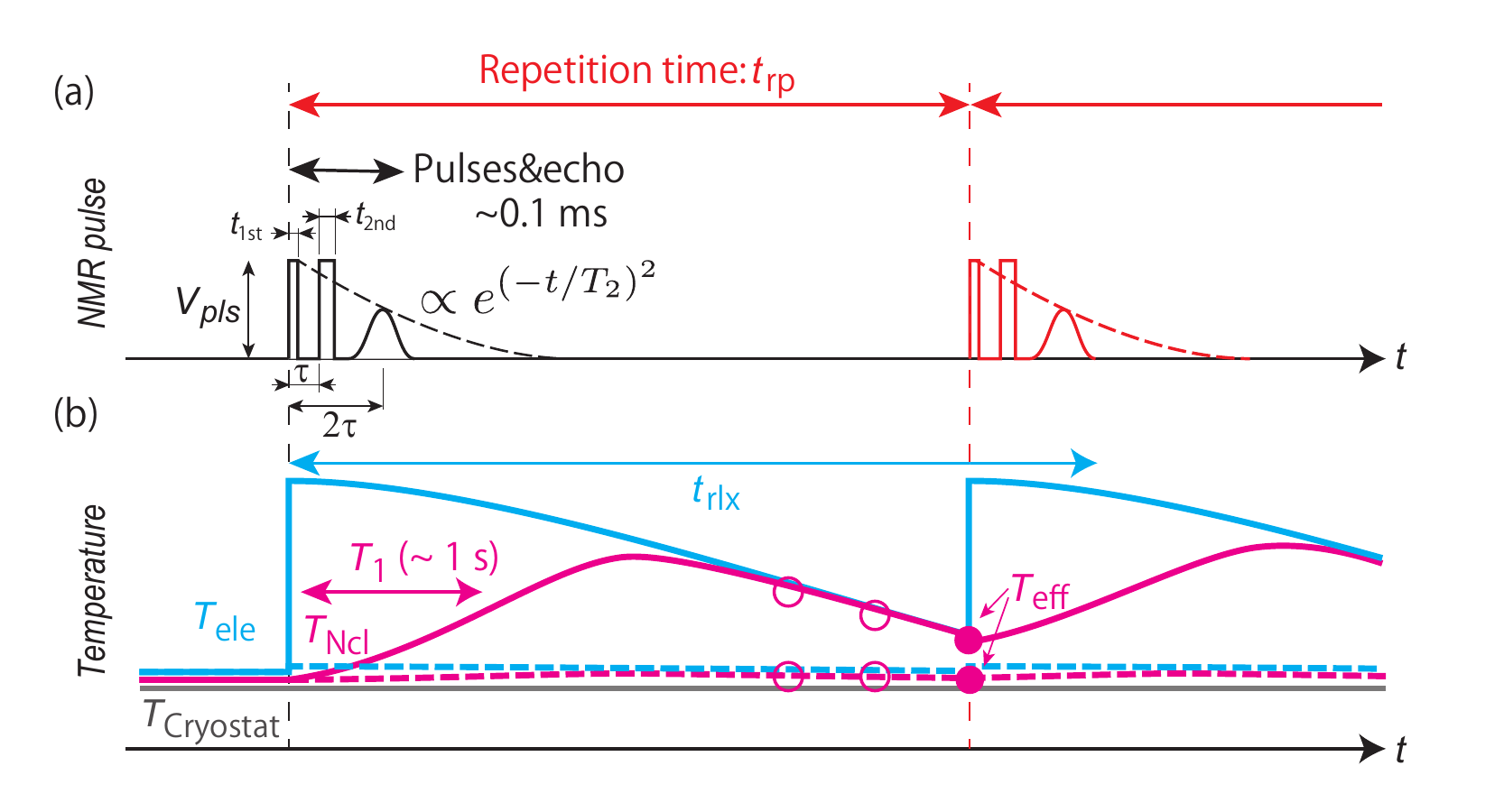}
	\caption{
		(a) An illustration of the NMR pulse sequence consisting of the first and second pulses of the width of $t_{\textrm{1st}}$ and $t_{\textrm{2nd}}$, and the amplitude $V_{\textrm{pls}}$. 
		A spin-echo signal is observed at $2\tau$ after the first pulse, where $\tau$ is the interval time between the first and the second pulses.
		This sequence is repeated with the repetition time ($t_{\textrm{rp}}$).
		The spin-echo decay by the transverse relaxation on the time scale of $T_2$ is illustrated by the dashed line.
		(b) The time dependences of various temperatures; the temperature of the cryostat ($T_{\textrm{Cryostat}}$, grey), the electrons ($T_{\textrm{ele}}$, blue) and the nuclear spins ($T_{\textrm{Ncl}}$, pink) are shown. The solid (dashed) lines show the case that the sample is heated up (not heated up) by the NMR pulse. Time scales of the longitudinal relaxation of the nuclear spins ($T_1$) and the thermal relaxation of the electron system ($t_{\textrm{rlx}}$) are also indicated by the arrows. Because of the repetition sequence, the effective temperature of the nuclear spins ($T_{\textrm{eff}}$, filled circles), where the NMR signals are accumulated, is affected by $t_{\textrm{rp}}$. Spin-echo signal is observed at a higher $T_{\textrm{eff}}$ (thus with a smaller spin-echo intensity) for a shorter $t_{\textrm{rp}}$ when the $T_{\textrm{ele}}$ is increased (open circles on the solid line). On the other hand, there is no repetition time dependence when the heating is negligible (open circles on the dashed line).
	}
	\label{PulseSeq}
\end{figure*}

To estimate the heating effect by checking the temperature dependence of $I_{\textrm{SE}}$, it is necessary to pay attention to the fact that the temperature of the electrons ($T_{\textrm{ele}}$) and that of the nuclear spins ($T_{\textrm{Ncl}}$) may be different in the sample in the time scale of the NMR measurements.
Figures\,\ref{PulseSeq} illustrate a schematic of the NMR pulse sequence (Fig.\,\ref{PulseSeq}(a)) and the time dependences of $T_{\textrm{ele}}$ and $T_{\textrm{Ncl}}$ (Fig.\,\ref{PulseSeq}(b)). The solid lines in Fig.\,\ref{PulseSeq}(b) show a case that the sample is heated up by an excess pulse power. In this case, the NMR pulses increase $T_{\textrm{ele}}$ which immediately affects the Knight shift. On the other hand, $T_{\textrm{Ncl}}$ remains the cryostat temperature ($T_{\textrm{Cryostat}}$) in the time scale of detecting the spin echo signal ($\ll T_1$), producing $I_{\textrm{SE}}$ at $T_{\textrm{Cryostat}}$.

For pulse NMR measurements, the pulse sequence is repeated at a repetition time ($t_{\textrm{rp}}$) interval in order to resolve a small signal by accumulating the NMR signals. Therefore, when $T_{\textrm{ele}}$ is increased by the NMR pulses with an excess $P_{\textrm{pls}}$, the heating of the sample increases the effective temperature of the nuclear spins ($T_{\textrm{eff}}$) where the NMR signals are accumulated. The effective temperature becomes higher (thus $I_{\textrm{SE}}$ becomes smaller) for a shorter $t_{\textrm{rp}}$ when the $T_{\textrm{ele}}$ is increased by the NMR pulses (the open circles on the solid line in Fig.\,\ref{PulseSeq}(b)). This repetition time dependence is absent when the increase of $T_{\textrm{ele}}$ is negligible (the open circles on the dashed line in Fig.\,\ref{PulseSeq}(b)). 

We checked this $t_{\textrm{rp}}$ dependence of $I_{\textrm{SE}}$ for $t_{\textrm{rp}}$ as short as possible because the heating effect becomes larger at a shorter $t_{\textrm{rp}}$ (see Fig.\,\ref{PulseSeq}(b)). 
As shown in Fig.\,\ref{PulseSeq}(b), the maximum of $T_{\textrm{Ncl}}$ is expected at $t \sim T_1$.
Therefore, the absence of the $t_{\textrm{rp}}$ dependence for $t_{\textrm{rp}} \sim T_1$ ensures the absence of the heating of the sample for a shorter time which may cover the time region required for $T_1$ measurements ($T_1/100 < t$). 
Given the higher sensitivity of $I_{\textrm{SE}}$ at lower temperatures (see Fig.\,\ref{NMRspct}(c)), this heating check procedure provides a sensitive method to confirm the sample temperature below 100\,mK.

\begin{figure}[!tbh]
	\centering
	\includegraphics[width=0.8\linewidth]{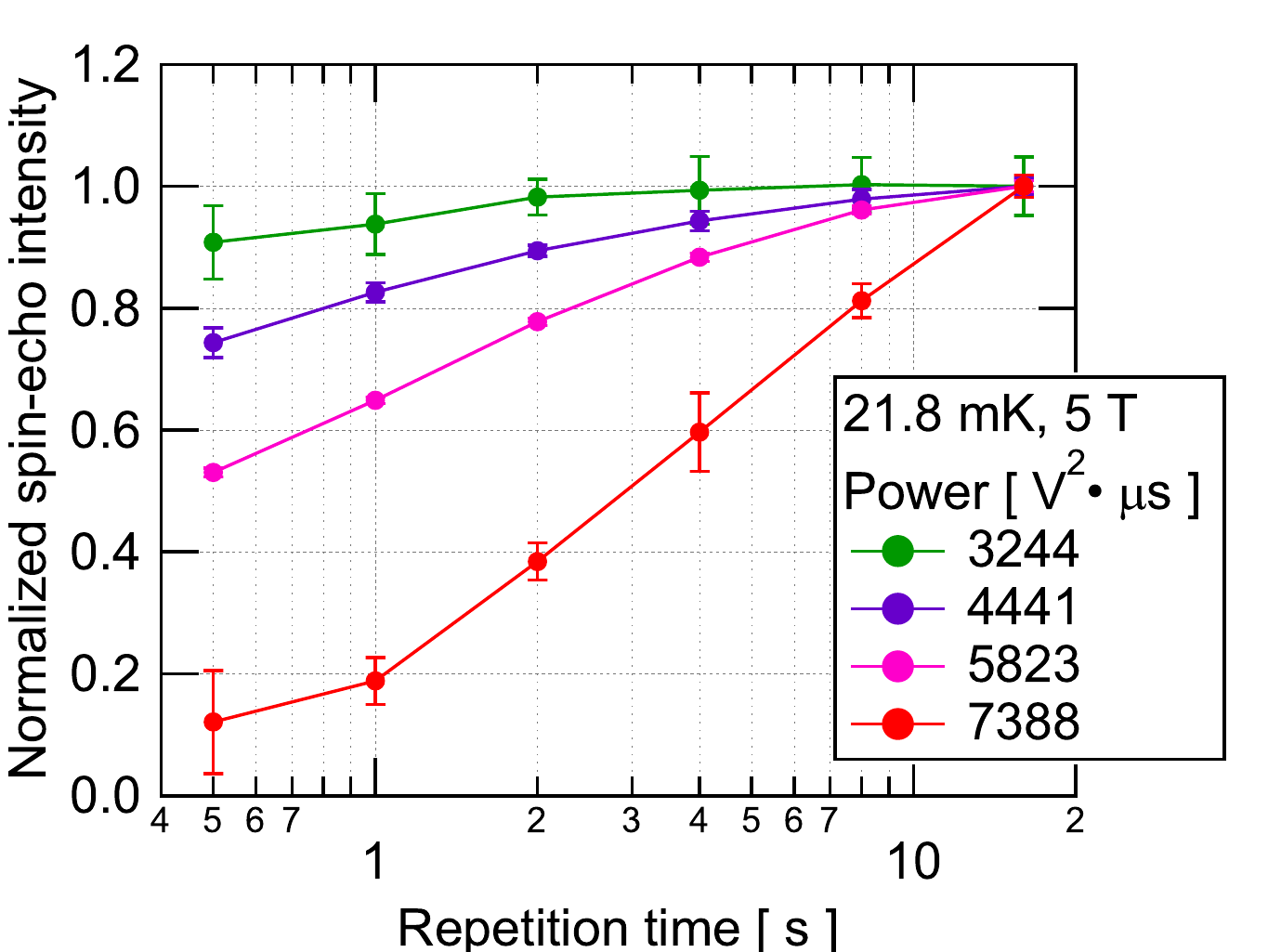}
	\caption{
		Repetition time dependence of the spin-echo intensity at 21.8\,mK and 5\,T. The data is normalized by the intensity at the longest repetition time. The signal becomes smaller at a shorter repetition time when a large pulse power is applied to the NMR circuit.
	}
	\label{RepDep}
\end{figure}

We first checked the $t_{\textrm{rp}}$ dependence of $I_{\textrm{SE}}$ for each temperature by using different pulse conditions. A typical dataset at 21.8 mK and 5 T is shown in Fig.\,\ref{RepDep}. As shown in Fig.\,\ref{RepDep}, smaller $I_{\textrm{SE}}$ was observed at a shorter $t_{\textrm{rp}}$ for a larger $P_{\textrm{pls}}$, showing the heating of the sample as expected in Fig.\,\ref{PulseSeq}. Such $t_{\textrm{rp}}$ dependence becomes smaller for smaller $P_{\textrm{pls}}$, and is almost absent at $P_{\textrm{pls}} = 3244$ V$^2\cdot \mu$s. We can thus find an appropriate pulse condition which does not heat up the sample by measuring the $t_{\textrm{rp}}$ dependence of $I_{\textrm{SE}}$.


\section{ Transverse relaxation time ($T_2$) measurements} \label{sec:T2}

To determine the temperature dependence of $I_{\textrm{SE}} (T)$, one needs to know the temperature dependence of $T_2$.
This is because the spin-echo signal, which is observed at a fixed delay time $2\tau$, decays as a function of $2\tau / T_2$ (see Fig.\,\ref{PulseSeq}(a)).
We thus checked the temperature dependence of $T_2$ by the pulse condition determined by the procedure described in Appendix\,\ref{sec:t_rp_dep}.

Figure \ref{T2}(a) shows the $\tau$ dependence of the spin-echo intensity of $^{59}$Co and $^{115}$In NMR at the In(2) site at 8\,T.
The decay curves can be fitted by a Gaussian function,  $I_{\textrm{SE}}(T, 2\tau)=I_{\textrm{SE}}(T) \exp \left\{ -(2\tau/T_2)^2\right\}$ for the whole temperature range we measured.
We find that $1/T_2$ decreases as lowering temperature as shown in Fig.\,\ref{T2}(b). This decrease might be explained by a decrease of the occupation number of the nuclear spins at $E_{1/2}$ and $E_{-1/2}$ for $k_B T < \gamma h H_0$.

\begin{figure}[!tbh]
	\centering
	\includegraphics[width=0.8\linewidth]{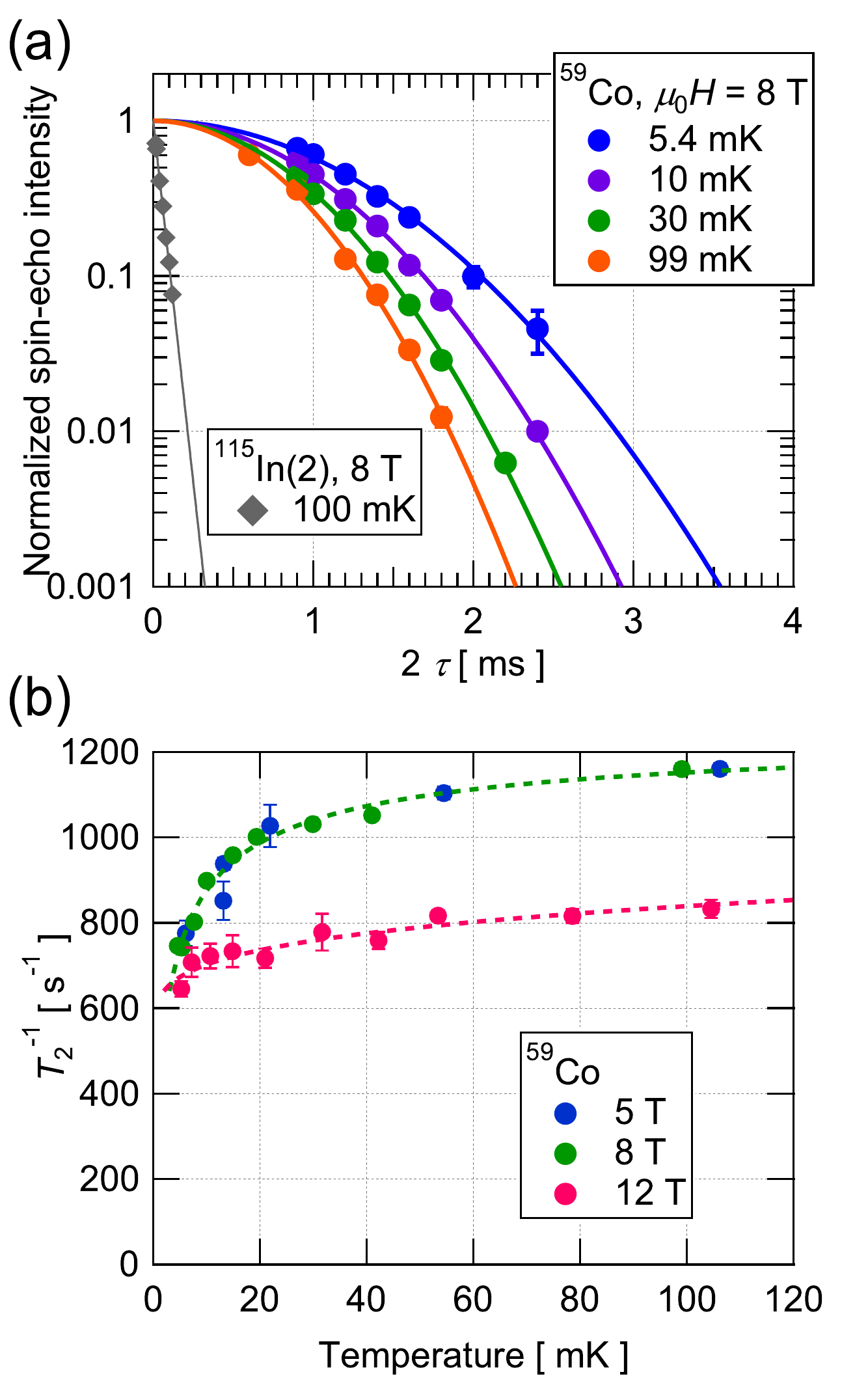}
	\caption{
		(a) The dependence of the spin-echo intensity on the interval time between the first and the second pulses ($\tau$) at different temperatures. The data is normalized by the value at $\tau=0$. The data of $^{59}$Co and $^{115}$In NMR at the In(2) site is shown by circles and diamonds, respectively. Each solid line shows a fit by $I_{\textrm{SE}}(T, 2\tau)=I_{\textrm{SE}}(T) \exp \left\{ -(2\tau/T_2)^2\right\}$. (b) The temperature dependence of $1/T_2$ of $^{59}$Co at 5, 8, and 12\,T. The dashed lines show a guide to the eyes.
	}
	\label{T2}
\end{figure}


\section{NMR intensity for $k_B T < \gamma h H_0$} \label{sec:I_SE_rel}

\begin{figure}[htb]
	\centering
	\includegraphics[width=0.8\linewidth]{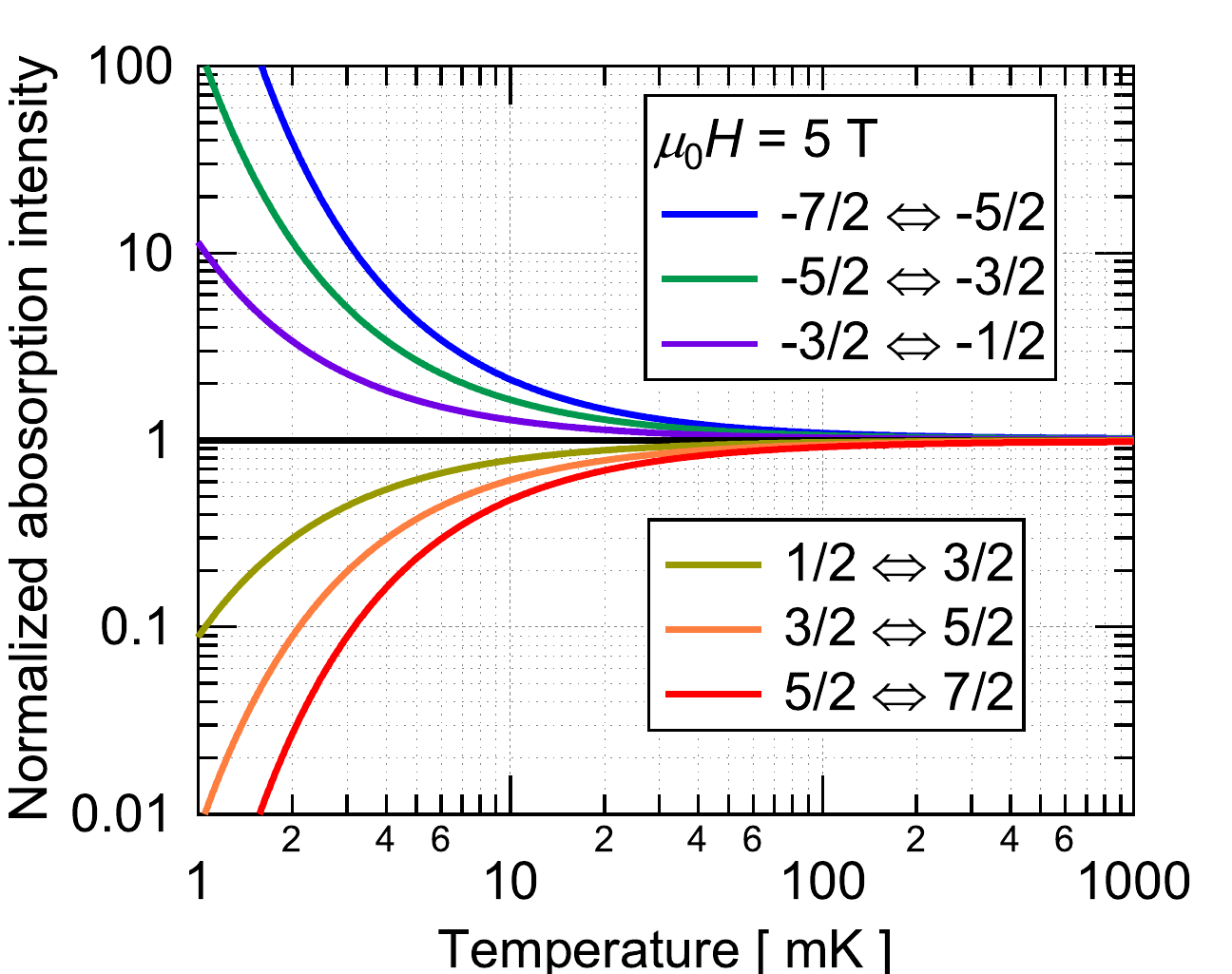}
	\caption{
		The temperature dependence of  $I_{\textrm{SE}}$ at 5\,T calculated for all NMR absorption in $^{59}$Co ($I = 7/2$) by using Eq.\,\ref{eq:I_SE}. The data is normalized by $I_{\textrm{SE}}$ of the center peak.
	}
	\label{I_SE_rel}
\end{figure}

As described in Section\,\ref{sec:verification}, the nuclear spins are redistributed to lower energy levels for $k_B T < \gamma h H_0$, giving rise to a larger NMR signals corresponding to the absorption between the lower energy levels.
To show this quantitatively, we calculate $I_{\textrm{SE}}$ at 5\,T by using Eq.\,\ref{eq:I_SE} for other transitions and plot the temperature dependence of $I_{\textrm{SE}}$ normalized by that of the center peak of $^{59}$Co ($m = +1/2 \leftrightarrow -1/2$) in Fig.\,\ref{I_SE_rel}.

As shown in Fig.\,\ref{I_SE_rel}, the relative NMR intensity for the absorption between the lower energy levels becomes larger for $k_B T < \gamma h H_0$, showing a new alternative possibility to extend the NMR measurements for lower temperatures.

\section{All $T_1$ data at 5\,T} \label{sec:T1table}

Since some of $1/T_1 T$ data points at 5\,T in Fig.\,\ref{T1} are overlapped each other, we list all the data in Table\,\ref{T1table}. The error represents one standard deviation of the data.

\begin{table}[htb]
	\caption{All data of the temperature dependence of $1/T_1 T$ at 5\,T shown in Fig.\,\ref{T1}.}
	\label{T1table}
	\begin{tabular}{ c  c } \toprule
		Temperature [ mK ] & $1/T_1 T$ [ sec$^{-1}$ K$^{-1} ]$\\ \midrule
	8.197 $\pm$ 0.034	& 13.77 $\pm$ 9.81 \\
	10.499 $\pm$ 0.085	& 17.84 $\pm$ 5.25 \\
	10.954 $\pm$ 0.034	& 18.28 $\pm$ 6.79 \\
	15.219 $\pm$ 0.030	& 57.37 $\pm$ 15.65 \\
	20.900 $\pm$ 0.032	& 66.16 $\pm$ 17.67 \\
	21.031 $\pm$ 0.035	& 70.21 $\pm$ 28.13 \\
	21.223 $\pm$ 0.027	& 79.41 $\pm$ 10.66 \\
	26.423 $\pm$ 0.028	& 43.66 $\pm$ 10.34 \\
	36.994 $\pm$ 0.038	& 53.27 $\pm$ 7.27 \\
	53.546 $\pm$ 0.052	& 57.58 $\pm$ 3.45 \\
	104.76 $\pm$ 0.082	& 44.18 $\pm$ 1.57 \\
	205.97 $\pm$ 0.12	& 35.68 $\pm$ 2.06 \\
	986.90 $\pm$ 0.13	& 28.42 $\pm$ 3.40 \\ \bottomrule
	\end{tabular}
\end{table}

\begin{acknowledgments}
This work was performed under the Visiting Researcher's Program of the Institute for Solid State Physics, University of Tokyo, and was supported by the Toray Science Foundation, and KAKENHI (Grants-in-Aid for Scientific Research) Grant Numbers 19H01848, 19K21842, and 19K23417.
\end{acknowledgments}

	
\newpage


\begin{thebibliography}{33}%
\makeatletter
\providecommand \@ifxundefined [1]{%
 \@ifx{#1\undefined}
}%
\providecommand \@ifnum [1]{%
 \ifnum #1\expandafter \@firstoftwo
 \else \expandafter \@secondoftwo
 \fi
}%
\providecommand \@ifx [1]{%
 \ifx #1\expandafter \@firstoftwo
 \else \expandafter \@secondoftwo
 \fi
}%
\providecommand \natexlab [1]{#1}%
\providecommand \enquote  [1]{``#1''}%
\providecommand \bibnamefont  [1]{#1}%
\providecommand \bibfnamefont [1]{#1}%
\providecommand \citenamefont [1]{#1}%
\providecommand \href@noop [0]{\@secondoftwo}%
\providecommand \href [0]{\begingroup \@sanitize@url \@href}%
\providecommand \@href[1]{\@@startlink{#1}\@@href}%
\providecommand \@@href[1]{\endgroup#1\@@endlink}%
\providecommand \@sanitize@url [0]{\catcode `\\12\catcode `\$12\catcode
  `\&12\catcode `\#12\catcode `\^12\catcode `\_12\catcode `\%12\relax}%
\providecommand \@@startlink[1]{}%
\providecommand \@@endlink[0]{}%
\providecommand \url  [0]{\begingroup\@sanitize@url \@url }%
\providecommand \@url [1]{\endgroup\@href {#1}{\urlprefix }}%
\providecommand \urlprefix  [0]{URL }%
\providecommand \Eprint [0]{\href }%
\providecommand \doibase [0]{https://doi.org/}%
\providecommand \selectlanguage [0]{\@gobble}%
\providecommand \bibinfo  [0]{\@secondoftwo}%
\providecommand \bibfield  [0]{\@secondoftwo}%
\providecommand \translation [1]{[#1]}%
\providecommand \BibitemOpen [0]{}%
\providecommand \bibitemStop [0]{}%
\providecommand \bibitemNoStop [0]{.\EOS\space}%
\providecommand \EOS [0]{\spacefactor3000\relax}%
\providecommand \BibitemShut  [1]{\csname bibitem#1\endcsname}%
\let\auto@bib@innerbib\@empty
\bibitem [{\citenamefont {Sachdev}(2010)}]{Sachdev2010}%
  \BibitemOpen
  \bibfield  {author} {\bibinfo {author} {\bibfnamefont {S.}~\bibnamefont
  {Sachdev}},\ }\href {https://doi.org/10.1002/pssb.200983037} {\bibfield
  {journal} {\bibinfo  {journal} {Phys. Status Solidi B}\ }\textbf {\bibinfo
  {volume} {247}},\ \bibinfo {pages} {537} (\bibinfo {year}
  {2010})}\BibitemShut {NoStop}%
\bibitem [{\citenamefont {Shibauchi}\ \emph {et~al.}(2014)\citenamefont
  {Shibauchi}, \citenamefont {Carrington},\ and\ \citenamefont
  {Matsuda}}]{Shibauchi2014}%
  \BibitemOpen
  \bibfield  {author} {\bibinfo {author} {\bibfnamefont {T.}~\bibnamefont
  {Shibauchi}}, \bibinfo {author} {\bibfnamefont {A.}~\bibnamefont
  {Carrington}},\ and\ \bibinfo {author} {\bibfnamefont {Y.}~\bibnamefont
  {Matsuda}},\ }\href
  {https://doi.org/10.1146/annurev-conmatphys-031113-133921} {\bibfield
  {journal} {\bibinfo  {journal} {Annual Review of Condensed Matter Physics}\
  }\textbf {\bibinfo {volume} {5}},\ \bibinfo {pages} {113} (\bibinfo {year}
  {2014})}\BibitemShut {NoStop}%
\bibitem [{\citenamefont {Gegenwart}\ \emph {et~al.}(2008)\citenamefont
  {Gegenwart}, \citenamefont {Si},\ and\ \citenamefont
  {Steglich}}]{Gegenwart2008}%
  \BibitemOpen
  \bibfield  {author} {\bibinfo {author} {\bibfnamefont {P.}~\bibnamefont
  {Gegenwart}}, \bibinfo {author} {\bibfnamefont {Q.}~\bibnamefont {Si}},\ and\
  \bibinfo {author} {\bibfnamefont {F.}~\bibnamefont {Steglich}},\ }\href
  {http://dx.doi.org/10.1038/nphys892} {\bibfield  {journal} {\bibinfo
  {journal} {Nature Physics}\ }\textbf {\bibinfo {volume} {4}},\ \bibinfo
  {pages} {186} (\bibinfo {year} {2008})}\BibitemShut {NoStop}%
\bibitem [{\citenamefont {Izawa}\ \emph {et~al.}(2001)\citenamefont {Izawa},
  \citenamefont {Yamaguchi}, \citenamefont {Matsuda}, \citenamefont {Shishido},
  \citenamefont {Settai},\ and\ \citenamefont {\=Onuki}}]{Izawa2001}%
  \BibitemOpen
  \bibfield  {author} {\bibinfo {author} {\bibfnamefont {K.}~\bibnamefont
  {Izawa}}, \bibinfo {author} {\bibfnamefont {H.}~\bibnamefont {Yamaguchi}},
  \bibinfo {author} {\bibfnamefont {Y.}~\bibnamefont {Matsuda}}, \bibinfo
  {author} {\bibfnamefont {H.}~\bibnamefont {Shishido}}, \bibinfo {author}
  {\bibfnamefont {R.}~\bibnamefont {Settai}},\ and\ \bibinfo {author}
  {\bibfnamefont {Y.}~\bibnamefont {\=Onuki}},\ }\href
  {https://doi.org/10.1103/PhysRevLett.87.057002} {\bibfield  {journal}
  {\bibinfo  {journal} {Phys. Rev. Lett.}\ }\textbf {\bibinfo {volume} {87}},\
  \bibinfo {pages} {057002} (\bibinfo {year} {2001})}\BibitemShut {NoStop}%
\bibitem [{\citenamefont {Kawasaki}\ \emph {et~al.}(2003)\citenamefont
  {Kawasaki}, \citenamefont {Kawasaki}, \citenamefont {Yashima}, \citenamefont
  {Mito}, \citenamefont {qing Zheng}, \citenamefont {Kitaoka}, \citenamefont
  {Shishido}, \citenamefont {Settai}, \citenamefont {Haga},\ and\ \citenamefont
  {\=Onuki}}]{Kawasaki2003}%
  \BibitemOpen
  \bibfield  {author} {\bibinfo {author} {\bibfnamefont {Y.}~\bibnamefont
  {Kawasaki}}, \bibinfo {author} {\bibfnamefont {S.}~\bibnamefont {Kawasaki}},
  \bibinfo {author} {\bibfnamefont {M.}~\bibnamefont {Yashima}}, \bibinfo
  {author} {\bibfnamefont {T.}~\bibnamefont {Mito}}, \bibinfo {author}
  {\bibfnamefont {G.}~\bibnamefont {qing Zheng}}, \bibinfo {author}
  {\bibfnamefont {Y.}~\bibnamefont {Kitaoka}}, \bibinfo {author} {\bibfnamefont
  {H.}~\bibnamefont {Shishido}}, \bibinfo {author} {\bibfnamefont
  {R.}~\bibnamefont {Settai}}, \bibinfo {author} {\bibfnamefont
  {Y.}~\bibnamefont {Haga}},\ and\ \bibinfo {author} {\bibfnamefont
  {Y.}~\bibnamefont {\=Onuki}},\ }\href {https://doi.org/10.1143/JPSJ.72.2308}
  {\bibfield  {journal} {\bibinfo  {journal} {Journal of the Physical Society
  of Japan}\ }\textbf {\bibinfo {volume} {72}},\ \bibinfo {pages} {2308}
  (\bibinfo {year} {2003})}\BibitemShut {NoStop}%
\bibitem [{\citenamefont {Tokiwa}\ \emph {et~al.}(2013)\citenamefont {Tokiwa},
  \citenamefont {Bauer},\ and\ \citenamefont {Gegenwart}}]{Tokiwa2013}%
  \BibitemOpen
  \bibfield  {author} {\bibinfo {author} {\bibfnamefont {Y.}~\bibnamefont
  {Tokiwa}}, \bibinfo {author} {\bibfnamefont {E.~D.}\ \bibnamefont {Bauer}},\
  and\ \bibinfo {author} {\bibfnamefont {P.}~\bibnamefont {Gegenwart}},\ }\href
  {https://doi.org/10.1103/PhysRevLett.111.107003} {\bibfield  {journal}
  {\bibinfo  {journal} {Phys. Rev. Lett.}\ }\textbf {\bibinfo {volume} {111}},\
  \bibinfo {pages} {107003} (\bibinfo {year} {2013})}\BibitemShut {NoStop}%
\bibitem [{\citenamefont {Bianchi}\ \emph {et~al.}(2003)\citenamefont
  {Bianchi}, \citenamefont {Movshovich}, \citenamefont {Vekhter}, \citenamefont
  {Pagliuso},\ and\ \citenamefont {Sarrao}}]{Bianchi2003}%
  \BibitemOpen
  \bibfield  {author} {\bibinfo {author} {\bibfnamefont {A.}~\bibnamefont
  {Bianchi}}, \bibinfo {author} {\bibfnamefont {R.}~\bibnamefont {Movshovich}},
  \bibinfo {author} {\bibfnamefont {I.}~\bibnamefont {Vekhter}}, \bibinfo
  {author} {\bibfnamefont {P.~G.}\ \bibnamefont {Pagliuso}},\ and\ \bibinfo
  {author} {\bibfnamefont {J.~L.}\ \bibnamefont {Sarrao}},\ }\href
  {https://doi.org/10.1103/PhysRevLett.91.257001} {\bibfield  {journal}
  {\bibinfo  {journal} {Phys. Rev. Lett.}\ }\textbf {\bibinfo {volume} {91}},\
  \bibinfo {pages} {257001} (\bibinfo {year} {2003})}\BibitemShut {NoStop}%
\bibitem [{\citenamefont {Paglione}\ \emph {et~al.}(2003)\citenamefont
  {Paglione}, \citenamefont {Tanatar}, \citenamefont {Hawthorn}, \citenamefont
  {Boaknin}, \citenamefont {Hill}, \citenamefont {Ronning}, \citenamefont
  {Sutherland}, \citenamefont {Taillefer}, \citenamefont {Petrovic},\ and\
  \citenamefont {Canfield}}]{Paglione2003}%
  \BibitemOpen
  \bibfield  {author} {\bibinfo {author} {\bibfnamefont {J.}~\bibnamefont
  {Paglione}}, \bibinfo {author} {\bibfnamefont {M.~A.}\ \bibnamefont
  {Tanatar}}, \bibinfo {author} {\bibfnamefont {D.~G.}\ \bibnamefont
  {Hawthorn}}, \bibinfo {author} {\bibfnamefont {E.}~\bibnamefont {Boaknin}},
  \bibinfo {author} {\bibfnamefont {R.~W.}\ \bibnamefont {Hill}}, \bibinfo
  {author} {\bibfnamefont {F.}~\bibnamefont {Ronning}}, \bibinfo {author}
  {\bibfnamefont {M.}~\bibnamefont {Sutherland}}, \bibinfo {author}
  {\bibfnamefont {L.}~\bibnamefont {Taillefer}}, \bibinfo {author}
  {\bibfnamefont {C.}~\bibnamefont {Petrovic}},\ and\ \bibinfo {author}
  {\bibfnamefont {P.~C.}\ \bibnamefont {Canfield}},\ }\href
  {https://doi.org/10.1103/PhysRevLett.91.246405} {\bibfield  {journal}
  {\bibinfo  {journal} {Phys. Rev. Lett.}\ }\textbf {\bibinfo {volume} {91}},\
  \bibinfo {pages} {246405} (\bibinfo {year} {2003})}\BibitemShut {NoStop}%
\bibitem [{\citenamefont {Howald}\ \emph {et~al.}(2011)\citenamefont {Howald},
  \citenamefont {Seyfarth}, \citenamefont {Knebel}, \citenamefont {Lapertot},
  \citenamefont {Aoki},\ and\ \citenamefont {Brison}}]{Howald2011}%
  \BibitemOpen
  \bibfield  {author} {\bibinfo {author} {\bibfnamefont {L.}~\bibnamefont
  {Howald}}, \bibinfo {author} {\bibfnamefont {G.}~\bibnamefont {Seyfarth}},
  \bibinfo {author} {\bibfnamefont {G.}~\bibnamefont {Knebel}}, \bibinfo
  {author} {\bibfnamefont {G.}~\bibnamefont {Lapertot}}, \bibinfo {author}
  {\bibfnamefont {D.}~\bibnamefont {Aoki}},\ and\ \bibinfo {author}
  {\bibfnamefont {J.-P.}\ \bibnamefont {Brison}},\ }\href
  {https://doi.org/10.1143/JPSJ.80.024710} {\bibfield  {journal} {\bibinfo
  {journal} {Journal of the Physical Society of Japan}\ }\textbf {\bibinfo
  {volume} {80}},\ \bibinfo {pages} {024710} (\bibinfo {year}
  {2011})}\BibitemShut {NoStop}%
\bibitem [{\citenamefont {Zaum}\ \emph {et~al.}(2011)\citenamefont {Zaum},
  \citenamefont {Grube}, \citenamefont {Sch\"afer}, \citenamefont {Bauer},
  \citenamefont {Thompson},\ and\ \citenamefont {v.~L\"ohneysen}}]{Zaum2011}%
  \BibitemOpen
  \bibfield  {author} {\bibinfo {author} {\bibfnamefont {S.}~\bibnamefont
  {Zaum}}, \bibinfo {author} {\bibfnamefont {K.}~\bibnamefont {Grube}},
  \bibinfo {author} {\bibfnamefont {R.}~\bibnamefont {Sch\"afer}}, \bibinfo
  {author} {\bibfnamefont {E.~D.}\ \bibnamefont {Bauer}}, \bibinfo {author}
  {\bibfnamefont {J.~D.}\ \bibnamefont {Thompson}},\ and\ \bibinfo {author}
  {\bibfnamefont {H.}~\bibnamefont {v.~L\"ohneysen}},\ }\href
  {https://doi.org/10.1103/PhysRevLett.106.087003} {\bibfield  {journal}
  {\bibinfo  {journal} {Phys. Rev. Lett.}\ }\textbf {\bibinfo {volume} {106}},\
  \bibinfo {pages} {087003} (\bibinfo {year} {2011})}\BibitemShut {NoStop}%
\bibitem [{\citenamefont {Sarrao}\ and\ \citenamefont
  {Thompson}(2007)}]{SarraoThompson2007}%
  \BibitemOpen
  \bibfield  {author} {\bibinfo {author} {\bibfnamefont {J.~L.}\ \bibnamefont
  {Sarrao}}\ and\ \bibinfo {author} {\bibfnamefont {J.~D.}\ \bibnamefont
  {Thompson}},\ }\href {https://doi.org/10.1143/JPSJ.76.051013} {\bibfield
  {journal} {\bibinfo  {journal} {Journal of the Physical Society of Japan}\
  }\textbf {\bibinfo {volume} {76}},\ \bibinfo {pages} {051013} (\bibinfo
  {year} {2007})}\BibitemShut {NoStop}%
\bibitem [{\citenamefont {Shishido}\ \emph {et~al.}(2018)\citenamefont
  {Shishido}, \citenamefont {Yamada}, \citenamefont {Sugii}, \citenamefont
  {Shimozawa}, \citenamefont {Yanase},\ and\ \citenamefont
  {Yamashita}}]{Shishido2018}%
  \BibitemOpen
  \bibfield  {author} {\bibinfo {author} {\bibfnamefont {H.}~\bibnamefont
  {Shishido}}, \bibinfo {author} {\bibfnamefont {S.}~\bibnamefont {Yamada}},
  \bibinfo {author} {\bibfnamefont {K.}~\bibnamefont {Sugii}}, \bibinfo
  {author} {\bibfnamefont {M.}~\bibnamefont {Shimozawa}}, \bibinfo {author}
  {\bibfnamefont {Y.}~\bibnamefont {Yanase}},\ and\ \bibinfo {author}
  {\bibfnamefont {M.}~\bibnamefont {Yamashita}},\ }\href
  {https://doi.org/10.1103/PhysRevLett.120.177201} {\bibfield  {journal}
  {\bibinfo  {journal} {Phys. Rev. Lett.}\ }\textbf {\bibinfo {volume} {120}},\
  \bibinfo {pages} {177201} (\bibinfo {year} {2018})}\BibitemShut {NoStop}%
\bibitem [{\citenamefont {Oja}\ and\ \citenamefont
  {Lounasmaa}(1997)}]{OjaLounasmaa1997}%
  \BibitemOpen
  \bibfield  {author} {\bibinfo {author} {\bibfnamefont {A.~S.}\ \bibnamefont
  {Oja}}\ and\ \bibinfo {author} {\bibfnamefont {O.~V.}\ \bibnamefont
  {Lounasmaa}},\ }\href {https://doi.org/10.1103/RevModPhys.69.1} {\bibfield
  {journal} {\bibinfo  {journal} {Rev. Mod. Phys.}\ }\textbf {\bibinfo {volume}
  {69}},\ \bibinfo {pages} {1} (\bibinfo {year} {1997})}\BibitemShut {NoStop}%
\bibitem [{\citenamefont {Buchal}\ \emph {et~al.}(1983)\citenamefont {Buchal},
  \citenamefont {Pobell}, \citenamefont {Mueller}, \citenamefont {Kubota},\
  and\ \citenamefont {Owers-Bradley}}]{Buchal1983}%
  \BibitemOpen
  \bibfield  {author} {\bibinfo {author} {\bibfnamefont {C.}~\bibnamefont
  {Buchal}}, \bibinfo {author} {\bibfnamefont {F.}~\bibnamefont {Pobell}},
  \bibinfo {author} {\bibfnamefont {R.~M.}\ \bibnamefont {Mueller}}, \bibinfo
  {author} {\bibfnamefont {M.}~\bibnamefont {Kubota}},\ and\ \bibinfo {author}
  {\bibfnamefont {J.~R.}\ \bibnamefont {Owers-Bradley}},\ }\href
  {https://doi.org/10.1103/PhysRevLett.50.64} {\bibfield  {journal} {\bibinfo
  {journal} {Phys. Rev. Lett.}\ }\textbf {\bibinfo {volume} {50}},\ \bibinfo
  {pages} {64} (\bibinfo {year} {1983})}\BibitemShut {NoStop}%
\bibitem [{\citenamefont {Tuoriniemi}\ \emph {et~al.}(2007)\citenamefont
  {Tuoriniemi}, \citenamefont {Juntunen-Nurmilaukas}, \citenamefont {Uusvuori},
  \citenamefont {Pentti}, \citenamefont {Salmela},\ and\ \citenamefont
  {Sebedash}}]{Tuoriniemi2007}%
  \BibitemOpen
  \bibfield  {author} {\bibinfo {author} {\bibfnamefont {J.}~\bibnamefont
  {Tuoriniemi}}, \bibinfo {author} {\bibfnamefont {K.}~\bibnamefont
  {Juntunen-Nurmilaukas}}, \bibinfo {author} {\bibfnamefont {J.}~\bibnamefont
  {Uusvuori}}, \bibinfo {author} {\bibfnamefont {E.}~\bibnamefont {Pentti}},
  \bibinfo {author} {\bibfnamefont {A.}~\bibnamefont {Salmela}},\ and\ \bibinfo
  {author} {\bibfnamefont {A.}~\bibnamefont {Sebedash}},\ }\href
  {https://doi.org/10.1038/nature05820} {\bibfield  {journal} {\bibinfo
  {journal} {Nature}\ }\textbf {\bibinfo {volume} {447}},\ \bibinfo {pages}
  {187} (\bibinfo {year} {2007})}\BibitemShut {NoStop}%
\bibitem [{\citenamefont {Schuberth}\ \emph {et~al.}(2016)\citenamefont
  {Schuberth}, \citenamefont {Tippmann}, \citenamefont {Steinke}, \citenamefont
  {Lausberg}, \citenamefont {Steppke}, \citenamefont {Brando}, \citenamefont
  {Krellner}, \citenamefont {Geibel}, \citenamefont {Yu}, \citenamefont {Si},\
  and\ \citenamefont {Steglich}}]{Schuberth2016}%
  \BibitemOpen
  \bibfield  {author} {\bibinfo {author} {\bibfnamefont {E.}~\bibnamefont
  {Schuberth}}, \bibinfo {author} {\bibfnamefont {M.}~\bibnamefont {Tippmann}},
  \bibinfo {author} {\bibfnamefont {L.}~\bibnamefont {Steinke}}, \bibinfo
  {author} {\bibfnamefont {S.}~\bibnamefont {Lausberg}}, \bibinfo {author}
  {\bibfnamefont {A.}~\bibnamefont {Steppke}}, \bibinfo {author} {\bibfnamefont
  {M.}~\bibnamefont {Brando}}, \bibinfo {author} {\bibfnamefont
  {C.}~\bibnamefont {Krellner}}, \bibinfo {author} {\bibfnamefont
  {C.}~\bibnamefont {Geibel}}, \bibinfo {author} {\bibfnamefont
  {R.}~\bibnamefont {Yu}}, \bibinfo {author} {\bibfnamefont {Q.}~\bibnamefont
  {Si}},\ and\ \bibinfo {author} {\bibfnamefont {F.}~\bibnamefont {Steglich}},\
  }\href {https://doi.org/10.1126/science.aaa9733} {\bibfield  {journal}
  {\bibinfo  {journal} {Science}\ }\textbf {\bibinfo {volume} {351}},\ \bibinfo
  {pages} {485} (\bibinfo {year} {2016})}\BibitemShut {NoStop}%
\bibitem [{\citenamefont {Prakash}\ \emph {et~al.}(2017)\citenamefont
  {Prakash}, \citenamefont {Kumar}, \citenamefont {Thamizhavel},\ and\
  \citenamefont {Ramakrishnan}}]{Prakash2017}%
  \BibitemOpen
  \bibfield  {author} {\bibinfo {author} {\bibfnamefont {O.}~\bibnamefont
  {Prakash}}, \bibinfo {author} {\bibfnamefont {A.}~\bibnamefont {Kumar}},
  \bibinfo {author} {\bibfnamefont {A.}~\bibnamefont {Thamizhavel}},\ and\
  \bibinfo {author} {\bibfnamefont {S.}~\bibnamefont {Ramakrishnan}},\ }\href
  {https://doi.org/10.1126/science.aaf8227} {\bibfield  {journal} {\bibinfo
  {journal} {Science}\ }\textbf {\bibinfo {volume} {355}},\ \bibinfo {pages}
  {52} (\bibinfo {year} {2017})}\BibitemShut {NoStop}%
\bibitem [{\citenamefont {Pustogow}\ \emph {et~al.}(2019)\citenamefont
  {Pustogow}, \citenamefont {Luo}, \citenamefont {Chronister}, \citenamefont
  {Su}, \citenamefont {Sokolov}, \citenamefont {Jerzembeck}, \citenamefont
  {Mackenzie}, \citenamefont {Hicks}, \citenamefont {Kikugawa}, \citenamefont
  {Raghu}, \citenamefont {Bauer},\ and\ \citenamefont {Brown}}]{Pustogow2019}%
  \BibitemOpen
  \bibfield  {author} {\bibinfo {author} {\bibfnamefont {A.}~\bibnamefont
  {Pustogow}}, \bibinfo {author} {\bibfnamefont {Y.}~\bibnamefont {Luo}},
  \bibinfo {author} {\bibfnamefont {A.}~\bibnamefont {Chronister}}, \bibinfo
  {author} {\bibfnamefont {Y.-S.}\ \bibnamefont {Su}}, \bibinfo {author}
  {\bibfnamefont {D.~A.}\ \bibnamefont {Sokolov}}, \bibinfo {author}
  {\bibfnamefont {F.}~\bibnamefont {Jerzembeck}}, \bibinfo {author}
  {\bibfnamefont {A.~P.}\ \bibnamefont {Mackenzie}}, \bibinfo {author}
  {\bibfnamefont {C.~W.}\ \bibnamefont {Hicks}}, \bibinfo {author}
  {\bibfnamefont {N.}~\bibnamefont {Kikugawa}}, \bibinfo {author}
  {\bibfnamefont {S.}~\bibnamefont {Raghu}}, \bibinfo {author} {\bibfnamefont
  {E.~D.}\ \bibnamefont {Bauer}},\ and\ \bibinfo {author} {\bibfnamefont
  {S.~E.}\ \bibnamefont {Brown}},\ }\href
  {https://doi.org/10.1038/s41586-019-1596-2} {\bibfield  {journal} {\bibinfo
  {journal} {Nature}\ }\textbf {\bibinfo {volume} {574}},\ \bibinfo {pages}
  {72} (\bibinfo {year} {2019})}\BibitemShut {NoStop}%
\bibitem [{\citenamefont {Ishida}\ \emph {et~al.}(2020)\citenamefont {Ishida},
  \citenamefont {Manago}, \citenamefont {Kinjo},\ and\ \citenamefont
  {Maeno}}]{Ishida2020}%
  \BibitemOpen
  \bibfield  {author} {\bibinfo {author} {\bibfnamefont {K.}~\bibnamefont
  {Ishida}}, \bibinfo {author} {\bibfnamefont {M.}~\bibnamefont {Manago}},
  \bibinfo {author} {\bibfnamefont {K.}~\bibnamefont {Kinjo}},\ and\ \bibinfo
  {author} {\bibfnamefont {Y.}~\bibnamefont {Maeno}},\ }\href
  {https://doi.org/10.7566/JPSJ.89.034712} {\bibfield  {journal} {\bibinfo
  {journal} {Journal of the Physical Society of Japan}\ }\textbf {\bibinfo
  {volume} {89}},\ \bibinfo {pages} {034712} (\bibinfo {year}
  {2020})}\BibitemShut {NoStop}%
\bibitem [{\citenamefont {Shishido}\ \emph {et~al.}(2002)\citenamefont
  {Shishido}, \citenamefont {Settai}, \citenamefont {Aoki}, \citenamefont
  {Ikeda}, \citenamefont {Nakawaki}, \citenamefont {Nakamura}, \citenamefont
  {Iizuka}, \citenamefont {Inada}, \citenamefont {Sugiyama}, \citenamefont
  {Takeuchi}, \citenamefont {Kindo}, \citenamefont {Kobayashi}, \citenamefont
  {Haga}, \citenamefont {Harima}, \citenamefont {Aoki}, \citenamefont {Namiki},
  \citenamefont {Sato},\ and\ \citenamefont {\=Onuki}}]{Shishido2002}%
  \BibitemOpen
  \bibfield  {author} {\bibinfo {author} {\bibfnamefont {H.}~\bibnamefont
  {Shishido}}, \bibinfo {author} {\bibfnamefont {R.}~\bibnamefont {Settai}},
  \bibinfo {author} {\bibfnamefont {D.}~\bibnamefont {Aoki}}, \bibinfo {author}
  {\bibfnamefont {S.}~\bibnamefont {Ikeda}}, \bibinfo {author} {\bibfnamefont
  {H.}~\bibnamefont {Nakawaki}}, \bibinfo {author} {\bibfnamefont
  {N.}~\bibnamefont {Nakamura}}, \bibinfo {author} {\bibfnamefont
  {T.}~\bibnamefont {Iizuka}}, \bibinfo {author} {\bibfnamefont
  {Y.}~\bibnamefont {Inada}}, \bibinfo {author} {\bibfnamefont
  {K.}~\bibnamefont {Sugiyama}}, \bibinfo {author} {\bibfnamefont
  {T.}~\bibnamefont {Takeuchi}}, \bibinfo {author} {\bibfnamefont
  {K.}~\bibnamefont {Kindo}}, \bibinfo {author} {\bibfnamefont {T.~C.}\
  \bibnamefont {Kobayashi}}, \bibinfo {author} {\bibfnamefont {Y.}~\bibnamefont
  {Haga}}, \bibinfo {author} {\bibfnamefont {H.}~\bibnamefont {Harima}},
  \bibinfo {author} {\bibfnamefont {Y.}~\bibnamefont {Aoki}}, \bibinfo {author}
  {\bibfnamefont {T.}~\bibnamefont {Namiki}}, \bibinfo {author} {\bibfnamefont
  {H.}~\bibnamefont {Sato}},\ and\ \bibinfo {author} {\bibfnamefont
  {Y.}~\bibnamefont {\=Onuki}},\ }\href {https://doi.org/10.1143/JPSJ.71.162}
  {\bibfield  {journal} {\bibinfo  {journal} {Journal of the Physical Society
  of Japan}\ }\textbf {\bibinfo {volume} {71}},\ \bibinfo {pages} {162}
  (\bibinfo {year} {2002})}\BibitemShut {NoStop}%
\bibitem [{\citenamefont {Kumagai}\ \emph {et~al.}(2011)\citenamefont
  {Kumagai}, \citenamefont {Shishido}, \citenamefont {Shibauchi},\ and\
  \citenamefont {Matsuda}}]{Kumagai2011}%
  \BibitemOpen
  \bibfield  {author} {\bibinfo {author} {\bibfnamefont {K.}~\bibnamefont
  {Kumagai}}, \bibinfo {author} {\bibfnamefont {H.}~\bibnamefont {Shishido}},
  \bibinfo {author} {\bibfnamefont {T.}~\bibnamefont {Shibauchi}},\ and\
  \bibinfo {author} {\bibfnamefont {Y.}~\bibnamefont {Matsuda}},\ }\href
  {https://doi.org/10.1103/PhysRevLett.106.137004} {\bibfield  {journal}
  {\bibinfo  {journal} {Phys. Rev. Lett.}\ }\textbf {\bibinfo {volume} {106}},\
  \bibinfo {pages} {137004} (\bibinfo {year} {2011})}\BibitemShut {NoStop}%
\bibitem [{\citenamefont {Sakai}\ \emph {et~al.}(2011)\citenamefont {Sakai},
  \citenamefont {Brown}, \citenamefont {Baek}, \citenamefont {Ronning},
  \citenamefont {Bauer},\ and\ \citenamefont {Thompson}}]{Sakai2011}%
  \BibitemOpen
  \bibfield  {author} {\bibinfo {author} {\bibfnamefont {H.}~\bibnamefont
  {Sakai}}, \bibinfo {author} {\bibfnamefont {S.~E.}\ \bibnamefont {Brown}},
  \bibinfo {author} {\bibfnamefont {S.-H.}\ \bibnamefont {Baek}}, \bibinfo
  {author} {\bibfnamefont {F.}~\bibnamefont {Ronning}}, \bibinfo {author}
  {\bibfnamefont {E.~D.}\ \bibnamefont {Bauer}},\ and\ \bibinfo {author}
  {\bibfnamefont {J.~D.}\ \bibnamefont {Thompson}},\ }\href
  {https://doi.org/10.1103/PhysRevLett.107.137001} {\bibfield  {journal}
  {\bibinfo  {journal} {Phys. Rev. Lett.}\ }\textbf {\bibinfo {volume} {107}},\
  \bibinfo {pages} {137001} (\bibinfo {year} {2011})}\BibitemShut {NoStop}%
\bibitem [{\citenamefont {Kohori}\ \emph {et~al.}(2001)\citenamefont {Kohori},
  \citenamefont {Yamato}, \citenamefont {Iwamoto}, \citenamefont {Kohara},
  \citenamefont {Bauer}, \citenamefont {Maple},\ and\ \citenamefont
  {Sarrao}}]{Kohori2001}%
  \BibitemOpen
  \bibfield  {author} {\bibinfo {author} {\bibfnamefont {Y.}~\bibnamefont
  {Kohori}}, \bibinfo {author} {\bibfnamefont {Y.}~\bibnamefont {Yamato}},
  \bibinfo {author} {\bibfnamefont {Y.}~\bibnamefont {Iwamoto}}, \bibinfo
  {author} {\bibfnamefont {T.}~\bibnamefont {Kohara}}, \bibinfo {author}
  {\bibfnamefont {E.~D.}\ \bibnamefont {Bauer}}, \bibinfo {author}
  {\bibfnamefont {M.~B.}\ \bibnamefont {Maple}},\ and\ \bibinfo {author}
  {\bibfnamefont {J.~L.}\ \bibnamefont {Sarrao}},\ }\href
  {https://doi.org/10.1103/PhysRevB.64.134526} {\bibfield  {journal} {\bibinfo
  {journal} {Phys. Rev. B}\ }\textbf {\bibinfo {volume} {64}},\ \bibinfo
  {pages} {134526} (\bibinfo {year} {2001})}\BibitemShut {NoStop}%
\bibitem [{\citenamefont {Curro}\ \emph {et~al.}(2001)\citenamefont {Curro},
  \citenamefont {Simovic}, \citenamefont {Hammel}, \citenamefont {Pagliuso},
  \citenamefont {Sarrao}, \citenamefont {Thompson},\ and\ \citenamefont
  {Martins}}]{Curro2001}%
  \BibitemOpen
  \bibfield  {author} {\bibinfo {author} {\bibfnamefont {N.~J.}\ \bibnamefont
  {Curro}}, \bibinfo {author} {\bibfnamefont {B.}~\bibnamefont {Simovic}},
  \bibinfo {author} {\bibfnamefont {P.~C.}\ \bibnamefont {Hammel}}, \bibinfo
  {author} {\bibfnamefont {P.~G.}\ \bibnamefont {Pagliuso}}, \bibinfo {author}
  {\bibfnamefont {J.~L.}\ \bibnamefont {Sarrao}}, \bibinfo {author}
  {\bibfnamefont {J.~D.}\ \bibnamefont {Thompson}},\ and\ \bibinfo {author}
  {\bibfnamefont {G.~B.}\ \bibnamefont {Martins}},\ }\href
  {https://doi.org/10.1103/PhysRevB.64.180514} {\bibfield  {journal} {\bibinfo
  {journal} {Phys. Rev. B}\ }\textbf {\bibinfo {volume} {64}},\ \bibinfo
  {pages} {180514(R)} (\bibinfo {year} {2001})}\BibitemShut {NoStop}%
\bibitem [{\citenamefont {Taniguchi}\ \emph {et~al.}(2020)\citenamefont
  {Taniguchi}, \citenamefont {Kitagawa}, \citenamefont {Manago}, \citenamefont
  {Nakamine}, \citenamefont {Ishida},\ and\ \citenamefont
  {Shishido}}]{Taniguchi2020}%
  \BibitemOpen
  \bibfield  {author} {\bibinfo {author} {\bibfnamefont {T.}~\bibnamefont
  {Taniguchi}}, \bibinfo {author} {\bibfnamefont {S.}~\bibnamefont {Kitagawa}},
  \bibinfo {author} {\bibfnamefont {M.}~\bibnamefont {Manago}}, \bibinfo
  {author} {\bibfnamefont {G.}~\bibnamefont {Nakamine}}, \bibinfo {author}
  {\bibfnamefont {K.}~\bibnamefont {Ishida}},\ and\ \bibinfo {author}
  {\bibfnamefont {H.}~\bibnamefont {Shishido}},\ }\bibfield  {booktitle} {\emph
  {\bibinfo {booktitle} {Proceedings of the International Conference on
  Strongly Correlated Electron Systems (SCES2019)}},\ }\href
  {https://doi.org/10.7566/JPSCP.30.011107} {\bibfield  {journal} {\bibinfo
  {journal} {JPS Conf. Proc.}\ }\textbf {\bibinfo {volume} {30}},\ \bibinfo
  {pages} {011107} (\bibinfo {year} {2020})}\BibitemShut {NoStop}%
\bibitem [{\citenamefont {Narath}(1967)}]{Narath1967}%
  \BibitemOpen
  \bibfield  {author} {\bibinfo {author} {\bibfnamefont {A.}~\bibnamefont
  {Narath}},\ }\href {https://doi.org/10.1103/PhysRev.162.320} {\bibfield
  {journal} {\bibinfo  {journal} {Phys. Rev.}\ }\textbf {\bibinfo {volume}
  {162}},\ \bibinfo {pages} {320} (\bibinfo {year} {1967})}\BibitemShut
  {NoStop}%
\bibitem [{foo()}]{footnote1}%
  \BibitemOpen
  \href@noop {} {}\bibinfo {note} {We note that our data above 100\,mK at 5\,T
  and 8\,T well reproduces the previous report of Ref.~\cite{Taniguchi2020} and
  Ref.~\cite{Sakai2011}, respectively. On the other hand, our data and that in
  Ref.~\cite{Taniguchi2020} at 5\,T deviate from that in Ref.~\cite{Sakai2011}
  at low temperatures. One possible reason for this deviation is a sample
  dependence; our measurements and those in Ref.~\cite{Taniguchi2020} were done
  in the crystal synthesized by the same researcher around the same time. Also,
  this deviation might be caused by a heating effect in Ref.~\cite{Sakai2011},
  because a heating of the sample results in an overestimation of $1/T_1 T$
  owing to a higher actual sample temperature.}\BibitemShut {Stop}%
\bibitem [{\citenamefont {Sakai}\ \emph {et~al.}(2010)\citenamefont {Sakai},
  \citenamefont {Baek}, \citenamefont {Brown}, \citenamefont {Ronning},
  \citenamefont {Bauer},\ and\ \citenamefont {Thompson}}]{Sakai2010}%
  \BibitemOpen
  \bibfield  {author} {\bibinfo {author} {\bibfnamefont {H.}~\bibnamefont
  {Sakai}}, \bibinfo {author} {\bibfnamefont {S.-H.}\ \bibnamefont {Baek}},
  \bibinfo {author} {\bibfnamefont {S.~E.}\ \bibnamefont {Brown}}, \bibinfo
  {author} {\bibfnamefont {F.}~\bibnamefont {Ronning}}, \bibinfo {author}
  {\bibfnamefont {E.~D.}\ \bibnamefont {Bauer}},\ and\ \bibinfo {author}
  {\bibfnamefont {J.~D.}\ \bibnamefont {Thompson}},\ }\href
  {https://doi.org/10.1103/PhysRevB.82.020501} {\bibfield  {journal} {\bibinfo
  {journal} {Phys. Rev. B}\ }\textbf {\bibinfo {volume} {82}},\ \bibinfo
  {pages} {020501} (\bibinfo {year} {2010})}\BibitemShut {NoStop}%
\bibitem [{\citenamefont {Bao}\ \emph {et~al.}(2000)\citenamefont {Bao},
  \citenamefont {Pagliuso}, \citenamefont {Sarrao}, \citenamefont {Thompson},
  \citenamefont {Fisk}, \citenamefont {Lynn},\ and\ \citenamefont
  {Erwin}}]{Bao2000}%
  \BibitemOpen
  \bibfield  {author} {\bibinfo {author} {\bibfnamefont {W.}~\bibnamefont
  {Bao}}, \bibinfo {author} {\bibfnamefont {P.~G.}\ \bibnamefont {Pagliuso}},
  \bibinfo {author} {\bibfnamefont {J.~L.}\ \bibnamefont {Sarrao}}, \bibinfo
  {author} {\bibfnamefont {J.~D.}\ \bibnamefont {Thompson}}, \bibinfo {author}
  {\bibfnamefont {Z.}~\bibnamefont {Fisk}}, \bibinfo {author} {\bibfnamefont
  {J.~W.}\ \bibnamefont {Lynn}},\ and\ \bibinfo {author} {\bibfnamefont
  {R.~W.}\ \bibnamefont {Erwin}},\ }\href
  {https://doi.org/10.1103/PhysRevB.62.R14621} {\bibfield  {journal} {\bibinfo
  {journal} {Phys. Rev. B}\ }\textbf {\bibinfo {volume} {62}},\ \bibinfo
  {pages} {R14621} (\bibinfo {year} {2000})}\BibitemShut {NoStop}%
\bibitem [{\citenamefont {Yashima}\ \emph {et~al.}(2009)\citenamefont
  {Yashima}, \citenamefont {Mukuda}, \citenamefont {Kitaoka}, \citenamefont
  {Shishido}, \citenamefont {Settai},\ and\ \citenamefont
  {\ifmmode~\bar{O}\else \={O}\fi{}nuki}}]{Yashima2009}%
  \BibitemOpen
  \bibfield  {author} {\bibinfo {author} {\bibfnamefont {M.}~\bibnamefont
  {Yashima}}, \bibinfo {author} {\bibfnamefont {H.}~\bibnamefont {Mukuda}},
  \bibinfo {author} {\bibfnamefont {Y.}~\bibnamefont {Kitaoka}}, \bibinfo
  {author} {\bibfnamefont {H.}~\bibnamefont {Shishido}}, \bibinfo {author}
  {\bibfnamefont {R.}~\bibnamefont {Settai}},\ and\ \bibinfo {author}
  {\bibfnamefont {Y.}~\bibnamefont {\ifmmode~\bar{O}\else \={O}\fi{}nuki}},\
  }\href {https://doi.org/10.1103/PhysRevB.79.214528} {\bibfield  {journal}
  {\bibinfo  {journal} {Phys. Rev. B}\ }\textbf {\bibinfo {volume} {79}},\
  \bibinfo {pages} {214528} (\bibinfo {year} {2009})}\BibitemShut {NoStop}%
\bibitem [{\citenamefont {Raymond}\ \emph {et~al.}(2007)\citenamefont
  {Raymond}, \citenamefont {Ressouche}, \citenamefont {Knebel}, \citenamefont
  {Aoki},\ and\ \citenamefont {Flouquet}}]{Raymond2007}%
  \BibitemOpen
  \bibfield  {author} {\bibinfo {author} {\bibfnamefont {S.}~\bibnamefont
  {Raymond}}, \bibinfo {author} {\bibfnamefont {E.}~\bibnamefont {Ressouche}},
  \bibinfo {author} {\bibfnamefont {G.}~\bibnamefont {Knebel}}, \bibinfo
  {author} {\bibfnamefont {D.}~\bibnamefont {Aoki}},\ and\ \bibinfo {author}
  {\bibfnamefont {J.}~\bibnamefont {Flouquet}},\ }\href
  {https://doi.org/10.1088/0953-8984/19/24/242204} {\bibfield  {journal}
  {\bibinfo  {journal} {Journal of Physics: Condensed Matter}\ }\textbf
  {\bibinfo {volume} {19}},\ \bibinfo {pages} {242204} (\bibinfo {year}
  {2007})}\BibitemShut {NoStop}%
\bibitem [{\citenamefont {Curro}\ \emph {et~al.}(2000)\citenamefont {Curro},
  \citenamefont {Hammel}, \citenamefont {Pagliuso}, \citenamefont {Sarrao},
  \citenamefont {Thompson},\ and\ \citenamefont {Fisk}}]{Curro2000}%
  \BibitemOpen
  \bibfield  {author} {\bibinfo {author} {\bibfnamefont {N.~J.}\ \bibnamefont
  {Curro}}, \bibinfo {author} {\bibfnamefont {P.~C.}\ \bibnamefont {Hammel}},
  \bibinfo {author} {\bibfnamefont {P.~G.}\ \bibnamefont {Pagliuso}}, \bibinfo
  {author} {\bibfnamefont {J.~L.}\ \bibnamefont {Sarrao}}, \bibinfo {author}
  {\bibfnamefont {J.~D.}\ \bibnamefont {Thompson}},\ and\ \bibinfo {author}
  {\bibfnamefont {Z.}~\bibnamefont {Fisk}},\ }\href
  {https://doi.org/10.1103/PhysRevB.62.R6100} {\bibfield  {journal} {\bibinfo
  {journal} {Phys. Rev. B}\ }\textbf {\bibinfo {volume} {62}},\ \bibinfo
  {pages} {R6100} (\bibinfo {year} {2000})}\BibitemShut {NoStop}%
\bibitem [{fn()}]{fn}%
  \BibitemOpen
  \href@noop {} {}\bibinfo {note} {It should be noted that a broadening of the
  NMR spectrum might be impossible to detect at $^{59}$Co NMR because the
  tetragonal symmetry of CeCoIn$_5$ may allow the magnetic moment to orient
  perpendicular to the in-plane component of the applied field. In fact, in the
  sister AFM compound CeRhIn$_5$, it has been confirmed that the magnetic
  moment points perpendicular to the in-plane magnetic field by the neutron
  scattering experiments~\cite{Raymond2007}.}\BibitemShut {Stop}%
\end{thebibliography}
\providecommand{\noopsort}[1]{}\providecommand{\singleletter}[1]{#1}%

\end{document}